\documentclass[preprint2,times]{aastex631}
\usepackage{multirow}
\usepackage{verbatim}
\usepackage{float}
\usepackage{graphicx}
\usepackage{subfigure}
\usepackage[section]{placeins}
\usepackage{xcolor}
\usepackage{ulem}
\newcommand{\hi}{{\rm H}{\sc i}}

\def\kms{\,km~s$^{-1}$}

\def\cm2{\,$\rm{cm^{-2}}$}
\def\cm3{\,$\rm{cm^{-3}}$}

\newcommand{\KIAA}{\affiliation{Kavli Institute for Astronomy and
Astrophysics, Peking University, 5 Yiheyuan Road, Haidian District, Beijing 100871, China}}
\newcommand{\DoA}{\affiliation{Department of Astronomy, School of Physics,
Peking University, 5 Yiheyuan Road, Haidian District, Beijing 100871, China}}

\begin{document}
\title{\large The Formation of Milky Way ``Bones'':\\
Ubiquitous HI Narrow Self-Absorption Associated with CO Emission}

\correspondingauthor{Ke Wang}
\email{kwang.astro@pku.edu.cn}
\author[0000-0002-9796-1507]{Shenglan Sun}\KIAA\DoA
\author[0000-0002-7237-3856]{Ke Wang}\KIAA
\author[0000-0001-8315-4248]{Xunchuan Liu}
\affiliation{Shanghai Astronomical Observatory, Chinese Academy of Sciences, Shanghai 200030, China}

\author[0000-0001-5950-1932]{Fengwei Xu}\KIAA\DoA

\begin{abstract}
Long and skinny molecular filaments running along Galactic spiral arms are known as ``bones'', since they make up the skeleton of the Milky Way. However, their origin is still an open question. Here, we compare spectral images of HI taken by FAST with archival CO and \emph{Herschel} dust emission to investigate the conversion from HI to H$_2$ in two typical Galactic bones, CFG028.68-0.28 and CFG047.06+0.26.
Sensitive FAST HI images and an improved methodology enabled us to extract HI narrow self-absorption (HINSA) features associated with CO line emission on and off the filaments, revealing the ubiquity of HINSA towards distant clouds for the first time. The derived cold HI abundances, [HI]/[H$_2$], of the two bones range from $\sim$(0.5 to 44.7)$\times10^{-3}$, which reveal different degrees of HI-H$_2$ conversion and are similar to that of nearby, low-mass star forming clouds, Planck Galactic cold clumps and a nearby active high-mass star forming region
G176.51+00.20.
The HI-H$_2$ conversion has been ongoing for 2.2 to 13.2 Myr in the bones, a timescale comparable to that of massive star formation therein. Therefore, we are witnessing young giant molecular clouds with rapid massive star formation. 
Our study paves the way of using HINSA to study cloud formation in Galactic bones, and more generally, in distant giant molecular clouds, in the FAST era.
\end{abstract}




\section{Introduction} \label{sec:introduction}
\par
High-mass star formation in the Milky Way, a barred spiral galaxy, mainly occurs in giant molecular clouds (GMCs) or giant molecular filaments \citep[GMFs,][]{Ragan2014,Abreu2016,ZhangMM2019}. A special type of GMFs, which is called the ``bone'' of the Milky Way, represents the densest structures associated with spiral arms \citep{2014ApJ...797...53G}. In recent years, more than 20 Milky Way bones have been discovered \citep{2015MNRAS.450.4043W,2016ApJS..226....9W,2018ApJ...864..153Z,2022ApJS..259...36G, Ge23, WangKe24_atlasFL}. 
The bones bridge Galactic spiral arms and local star formation activities \citep{2015MNRAS.450.4043W}, and present some of the fundamental kinematic properties of GMCs in general \citep{WangKe24_atlasFL}.
Therefore, it is of great importance to investigate how these structures form and what role they play in star formation.

The formation of Galactic bones, and more general GMCs, lies in the chemical balance between atomic hydrogen (HI) and molecular hydrogen ($\mathrm{H}_{2}$). On one hand, HI is shielded from the interstellar UV radiation in the central, dense region of molecular clouds and cools to form $\mathrm{H}_{2}$ on grain surfaces. On the other hand, HI is also produced by cosmic ray dissociation of $\mathrm{H}_{2}$. So, the ratio [HI]/[H$_2$], which encodes how many cold H atoms are transformed into $\mathrm{H}_{2}$ is an indicator of bone ages. 

\par
HI Self-Absorption (HISA) occurs when dense, cold atomic hydrogen is positioned in front of a warmer emission background, resulting in an absorption dip that can be used to trace cold atomic gas \citep{1974AJ.....79..527K,2010ApJ...724.1402K}. HI narrow self-absorption (HINSA) is a specific type of HISA that has been observed in nearby molecular clouds \citep[e.g.,][]{2005ApJ...622..938G,2008ApJ...689..276K,2018ApJ...867...13Z}. Unlike HISA, HINSA exhibits a linewidth comparable to CO \citep{2003ApJ...585..823L} and shows coherent velocity and spatial structures with molecular components \citep{2003ApJ...585..823L,2005ApJ...622..938G}. Consequently, the atomic gas traced by HINSA is likely more closely associated with the processes responsible for molecular synthesis within the clouds. In recent years, HINSA has revealed the conversion from cold HI to H$_2$ in nearby molecular clouds. For example, \cite{2018ApJ...867...13Z} identified a ``ring'' of enhanced HI abundance surrounding a forming cloud B227. HINSA is also observed towards Planck Galactic cold clumps \citep[PGCCs;][]{2016A&A...594A..28P}, with a detection rate ranging from $\sim$58\% \citep{2020RAA....20...77T} to $\sim$92\% \citep{2022A&A...658A.140L}. PGCCs are mostly quiescent \citep{2012ApJ...756...76W}, probing the initial stages of star formation. 

To date, there are only a few cases of HISA observations towards high-mass star forming regions \citep{2020A&A...634A.139W,2020A&A...638A..44B,2022ApJ...933L..26L,LiC2023_CygX_HI}, simply because of complicated HI velocity components along the line of sight. Leveraging the high spectral resolution and high sensitivity provided by Five-hundred-meter Aperture Spherical radio Telescope \citep[FAST;][]{2011IJMPD..20..989N}, we can now explore the transition from atomic to molecular hydrogen by HINSA in two distant Milky Way bones, ``cold filament'' CFG028.68-0.28 (hereafter G28) and CFG047.06+0.26 (hereafter G47). The paper is organized as follows. The sample of two Milky Way bones are listed in Section \ref{sec:sample}. The FAST observations, data reduction, and retrieved archival data are summarized in Section \ref{sec:observation}. The methods and results are shown in Section \ref{sec:analysis}. The interpretation about the cloud age are discussed in Section \ref{sec:discussion}. Lastly, we make a summary in Section \ref{sec:summary}.

\section{Two Milky Way Bones} \label{sec:sample}
G28 and G47 were identified and characterized by \cite{2015MNRAS.450.4043W} in \textit{Herschel} images. The ``cold filament'' G28 is an S-shaped structure of 60 pc long running along the Scutum arm. G47 is an L-shaped filament of 78 pc running along the Sagittarius arm. At a distance of 4.89/4.44 kpc \citep{2015MNRAS.450.4043W}, G28/G47 have molecular gas mass of 5.5/2.0$\times 10^4 M_\odot$, representative of Galactic bones and more general giant molecular filaments. G28 and G47 are overall cold, with dust temperature in range of 17-32 K, and are associated with IRDCs \citep{2015ApJ...805..171L}. They both have potential of forming massive star clusters, and some star formation activity has been reported in both of them \citep{2018A&A...609A..43X,2019RAA....19..183X}.

G47 has been also observed as part of the SOFIA legacy project FIELDMAPS to map 10 IR-dark bones in 214 $\mu$m polarized dust emission \citep{2022ApJ...926L...6S}. The ordered magnetic field lines perpendicular to the spine of the G47, suggesting that B-fields likely play a role in supporting the G47 from collapse. \cite{2022Natur.601...49C} successfully detected Zeeman effect of HINSA, and derived magnetic field strength of L1544, a low-mass prestellar core in the nearby Taurus molecular cloud using FAST. If a large number of deep and narrow HINSA signatures can be identified in giant molecular clouds, we can expect to measure magnetic field strength routinely with FAST.

\section{Observations and Data}\label{sec:observation}
\subsection{FAST HI Mapping}
HI spectral mapping observations were conducted using the FAST on 2022 April 22 and 24 for G28 and G47, respectively. \footnote{Earlier observations were taken on 2021 November 17 and December 6, but  the absolute flux cannot be calibrated due to an error in observing scripts. In this study we use the the 2022 April observations.} The observations utilized the 19-beam receiver in the MultiBeamOTF mode. On-the-fly (OTF) maps are obtained by scanning along RA and DEC, respectively. The map size is $0^{\circ}.5\times 0^{\circ}.5$
centered on (RA, Dec) = (19:16:25.07, +12:44:06) at Epoch of J2000.

Following \cite{2019SCPMA..6259508Q}, the 19-beam receiver's footprint was rotated by $23^{\circ}.413$ with respect to the scanning direction, so that the 19 scanning stripes are evenly separated by $1'.17$, resulting an over Nyquist sampling. The scanning speed is $6''$/s, or equivalently 117\,s per beam, after combining one RA and one DEC map. For flux calibration, a 10\,K modulated noise is tuned on for 1\,s and off for 1\,s, repeatedly throughout the observations. Data is recorded every 1\,s.

The SPEC(W+N) backend is used to simultaneously record wide band (500 MHz) and narrow band (31.25\,MHz) spectra in full polarization. With 64\,k channels, the channel width is 1.5 and 0.1\kms\ for the wide and narrow band, respectively. Here we use only the narrow band data.

The pointing accuracy of FAST is better than $16''$ \citep{2019SCPMA..6259502J}. The FAST has an illuminated diameter of 300 m, and its beam is $2'.9$ at the HI line frequency of 1420.405 GHz. The rms noise measures $\sim$6 mK per 0.1\kms\ channel width.

G28 was observed in a similar way as G47, with difference in scanning speed ($15''$/s), noise modulation (1 s on followed by 31 s off), and coordinates ($2^{\circ}\times 2^{\circ}$ centered on RA, Dec = 18:43:31.59, -03:59:30.5). The resulting rms is $\sim$10 mK per 0.1\kms\ channel.

\subsection{FAST HI Data Reduction}

We use HiFAST \citep{JingY24_HiFAST} to reduce the data.  
The first step is flux calibration. The program calculates $T_{a}\left ( \nu  \right )  $ by dividing and comparing the power values when the periodically injected noise diode is on and off \citep{2020RAA....20...64J}. 
Secondly, sky coordinates are calculated using the positioning of the feed cabin.
Thirdly, gain values of the 19 beams \citep{2020RAA....20...64J,2020Innov...100053Q} are used to calibrate out the slightly different beam-to-beam response. 
After removing baseline and negligible radio frequency interference (RFI), a final data cube is formed by combining orthogonal scans (along RA and Dec).


\subsection{CO and Dust Data}

We retrieved $^{13}$CO $\left ( J= 1-0  \right ) $ data cubes of G28 and G47 from the Galactic Ring Survey \citep{2006ApJS..163..145J}. The angular resolution is 46$^{\prime \prime}$, and the rms noise is 0.4 K per $0.2 \mathrm{~km} \mathrm{~s}^{-1}$ channel width.
The $^{12}$CO $\left ( J= 1-0  \right ) $ cubes of G28 and G47 are obtained from the FOREST Unbiased Galactic-plane Imaging survey with
the Nobeyama 45-m telescope (FUGIN; \cite{2017PASJ...69...78U}) with angular resolution of 20$^{\prime \prime}$ for $^{12}$CO. The average rms noise levels in main beam temperature b scale with a velocity resolution of 1.3 km s$^{-1}$ were $\sim$1.47 K for $^{12}$CO with $8.^{\prime \prime}5$ pixel.


The H$_{2}$ column density and dust temperature data of G28 and G47 are taken from \cite{marsh2017}. These are computed by fitting SED built from Herschel images, with a resolution of $28''$.


\section{HINSA Analysis and Results}\label{sec:analysis}
\par
According to \cite{2008ApJ...689..276K}, a small HINSA dip can lead to a dominant feature in the second derivative representation of the observed HI spectrum, so it is sufficient to extract the HINSA feature by minimizing the integrated squared sum of the second derivative background spectrum. This method estimates central velocity, line width and temperature of cold HI gas from molecular tracers, and adopts trial values of optical depth to recover the background HI spectrum T$_{\mathrm{HI}}$ from observed spectrum T$_{\mathrm{r}}$. The values that minimize the squared integrated intensity of the second derivative of T$_{\mathrm{HI}}$ are the best-fit values of HINSA parameters.
\par
We extract HINSA features following the method in Section 4.2 of \cite{2022A&A...658A.140L}, which is an improvement of the method in \cite{2008ApJ...689..276K}. The improved method has much lower SNR requirement and performs better for large line width resultant from blended velocity components or high velocity resolution.

\begin{figure*}[htbp]
\epsscale{1.3}
\plotone{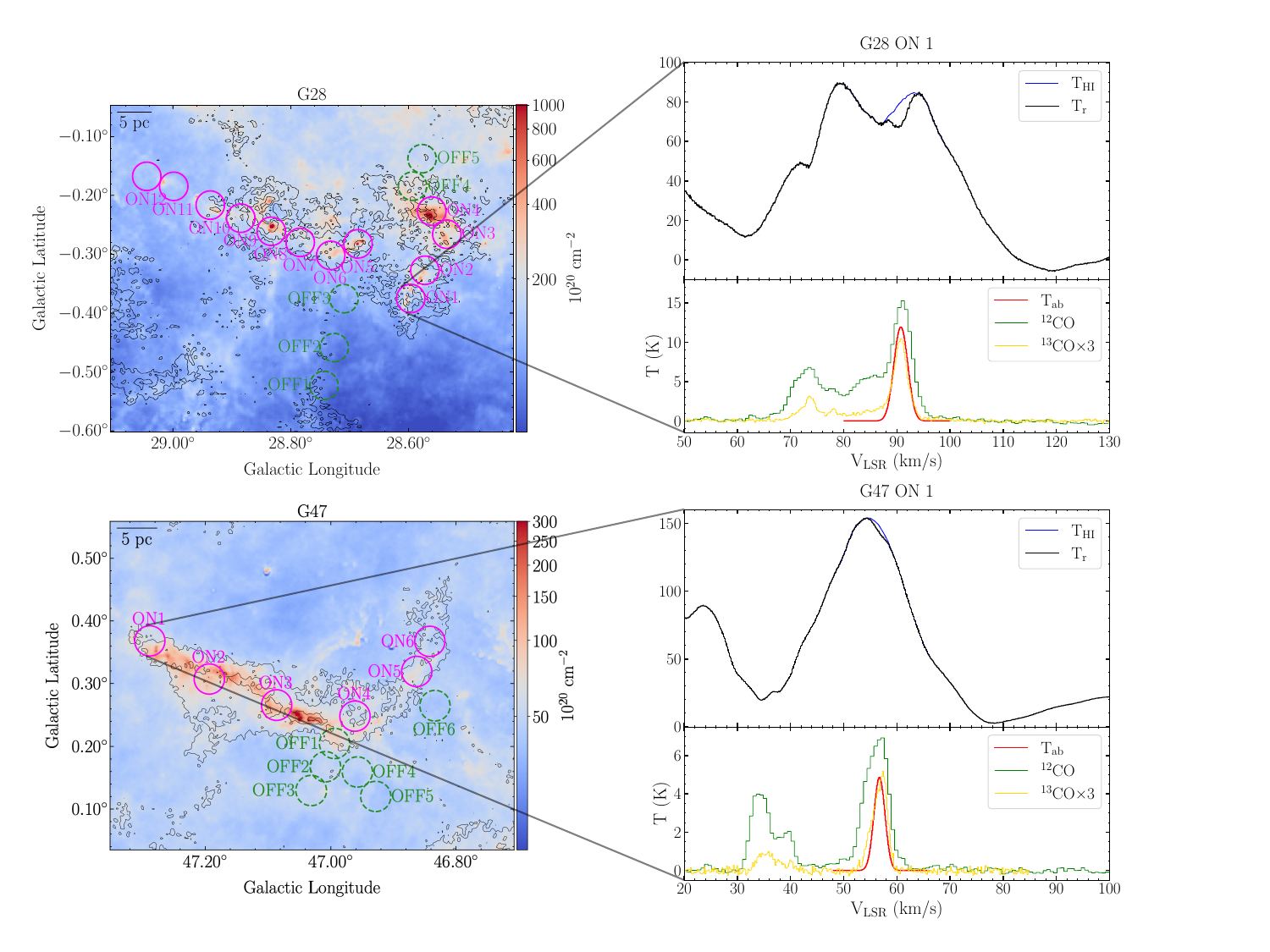}
\caption{%
Left: $^{12}$CO integrated intensity contours overlaid on H$_2$ column density from \cite{marsh2017} for G28 (top) and G47 (bottom). The contours are at 6, 9, 12$\times 10^{4}$K km s$^{-1}$ and 4, 8, 12$\times 10^{4}$K km s$^{-1}$, respectively. The magenta solid circles are ON-filament regions and the green dashed circles are OFF-filament regions, both with a diameter of FAST beam ($\sim2'.9$). Right: HINSA fitting and CO spectra towards representative regions of G28 (top, region ON 1) and G47 (bottom, region ON 1). For each region, recovered HI background T$_{\mathrm{HI}}$ (blue) and observed HI spectra T$_{\mathrm{r}}$ (black) are in the top half. Fitted spectrum of HINSA features T$_{\mathrm{ab}}$=T$_{\mathrm{HI}}-$T$_{\mathrm{r}}$ (red), $^{12}$CO (green) and  $^{13}$CO (gold) are in the second half. The temperatures of $^{13}$CO are magnified 3 times.} 
\label{fig:4Figs}
\end{figure*}

\par
In order to reduce the influence of observed HI spectral line jitters on fitting results, a moving averages of four adjacent channels of observed HI spectra are performed before fitting.
We search for HINSA features primarily according to its corresponding $^{13}$CO and $^{12}$CO emission. From a series of close circular regions (with the same size as FAST beam) along the spines of G28 and G47, we find HINSA in several ``ON-filament'' regions. In the regions that lie along spines but are not selected as ON regions, there are CO emission peaks without corresponding HI dips or vice versa in the velocity range of each GMF. Like the ``ring'' of enhanced HI abundance surrounding the  evolving molecular cloud B227 in \cite{2018ApJ...867...13Z}, we also select several ``OFF-filament'' regions with HINSA features around these two filaments to construct similar rings to observe HI abundance variation around the filaments.
Note that we only search for HINSA in the ranges of radial velocity corresponding to G28 and G47 reported in \cite{2015MNRAS.450.4043W}; other velocity ranges may well have HINSA features along the line-of-sight, but are not associated with the bones of interest in this study. See Figure\,\ref{fig:4Figs} for examples.

\par
The initial values of the free parameters, i.e. the FWHM linewidth $\Delta V$ and the centroid velocity $V_{\mathrm{LSR}}$ are estimated from spectral lines of $^{13}$CO or $^{12}$CO. Excitation temperatures of the cold HI components $T_{\mathrm{ex}}$ are assumed to be 10 K. Then the best-fit values of the optical depth $\tau_{0}$, $\Delta V$ and $V_{\mathrm{LSR}}$ are derived and the background spectrum $T_{\mathrm{HI}}$ is recovered. If two CO components are observed to correspond to two HI dips, then the fitting model is adjusted to double HINSA component. Examples of HINSA fitting are in Figure\,\ref{fig:4Figs}. As shown in Figure\,\ref{fig:4Figs}, HINSA features are found in 17 regions and 12 regions towards G28 and G47, respectively (see Appendix\,\ref{Appendix}). All HINSA features are coherent with CO lines in position-position-velocity (PPV) space so that the features trace the atomic-to-molecular transition in G28 and G47.

\par
The fitting results of $\tau_{0}$, $\Delta V$, $V_{\mathrm{LSR}}$, and the peak intensity, $T_{\mathrm{ab}}$, with 1 $\sigma$ errors are listed in Table\,\ref{table1}. The column density of cold HI traced by HINSA can be calculated as \citep{2003ApJ...585..823L}
\begin{equation}
    N\left(\mathrm{HINSA}\right)=1.9 \times 10^{18} \tau_{0} \frac{T_{\mathrm{ex}}}{\mathrm{K}} \frac{\Delta V}{\mathrm{~km} \mathrm{~s}^{-1}} \mathrm{~cm}^{-2},
\end{equation}
where $\Delta V=\sqrt{8 \ln (2)} \sigma_{\mathrm{HINSA}}$. $\sigma_{\mathrm{HINSA}}$ is the velocity dispersion of HINSA. 
According to Eq. 12 in \cite{2003ApJ...585..823L}, $\tau_0$ can be semiquantitatively corrected to $\tau_0/p$, in which $\tau_b=p(\tau_f+\tau_b)$, same as \cite{2003ApJ...585..823L}. $\tau_f$ and $\tau_b$ are the foreground and background HI optical depth. The method for estimating the p-values of these two GMFs is as follows: define point A as the intersection of the Sun's orbit around the Galactic center with the line of sight direction of each GMF. The p-value is then obtained by dividing the distance from each GMF to point A by the distance from the Sun to point A. $p\approx0.65$ and $p\approx0.59$ are derived for G28 and G47.
The corrected $N$(HINSA) of each HINSA component and cold HI abundance, [HI]/[H$_2$]=$N\left(\mathrm{HINSA}\right)/N_{\mathrm{H}_2}$, of each region are listed in Table\,\ref{table1}. $N_{\mathrm{H}_2}$ is the column density of H$_2$. In ON regions of G28 and G47, $N$(HINSA) ranges from $\sim$(0.4 to 14.1)$\times10^{19}$ cm$^{-2}$ and [HI]/[H$_2$] ranges from $\sim$(0.5 to 7.2)$\times10^{-3}$. In all OFF regions, $N$(HINSA) ranges from $\sim$(1.8 to 32.8)$\times10^{19}$ cm$^{-2}$ and [HI]/[H$_2$] ranges from $\sim$(1.9 to 44.7)$\times10^{-3}$.
\par
The FWHM linewidth and central velocity of each $^{13}$CO component corresponding to HINSA are fitted using Gaussian model and listed in Table\,\ref{table1}, too, in which the FWHM of each $^{13}$CO component $\Delta V(^{13}\mathrm{CO})=\sqrt{8 \ln (2)} \sigma_{^{13}\mathrm{CO}}$. $\sigma_{^{13}\mathrm{CO}}$ is the velocity dispersion of the $^{13}$CO component. For G47 OFF 6, two HINSA components corresponds to only one $^{13}$CO peak.

\section{Discussion}\label{sec:discussion}
\subsection{Cloud Ages}
\begin{figure*}[htbp]
\epsscale{0.8}
\plotone{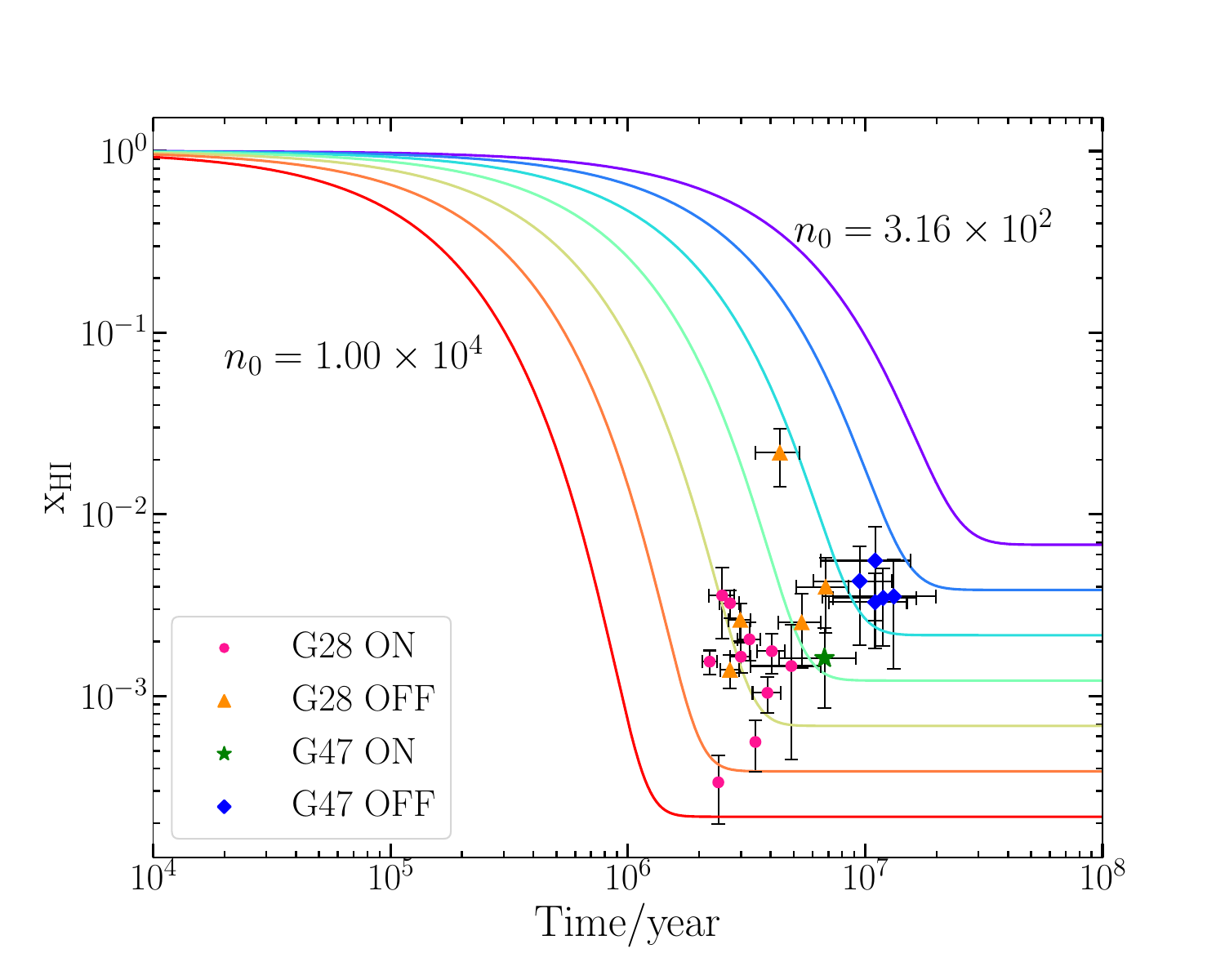}
\caption{Time dependence of $x_{\mathrm{HI}}$ for different total proton density (in cm$^{-3}$). The values of total proton densities of the colorful curves are logarithmically spaced at intervals of 0.25, from $10^{2.50}=3.16\times10^2$ to $10^4$. 
G28 ON, G28 OFF, G47 ON and G47 OFF regions are plotted with pink dots, orange triangles, green star and blue diamonds, respectively.
The fractional abundances of cold HI of G28 ON 3 and ON 4 are very close to the steady state value so their errorbars of time are not plotted.}
\label{fig:chemical}
\end{figure*}
We compare our results of abundance ratio [HI]/[H$_{2}$] to the analytical model proposed by \cite{2005ApJ...622..938G} to estimate the timescale for evolution of G28 and G47. According to the model, the fractional abundance of cold HI varies with time as,
\begin{equation}\label{eq2}
    x_{\mathrm{HI}}(t)=1-\frac{2 k^{\prime} n_{0}}{2 k^{\prime} n_{0}+\zeta_{\mathrm{H}_{2}}}\left[1-\exp \left(\frac{-t}{\tau_{\mathrm{HI} \rightarrow \mathrm{H}_{2}}}\right)\right],
\end{equation}
in which $k^{\prime}$ is the formation rate coefficient. $\zeta_{\mathrm{H}_{2}}$ is the ionization rate by cosmic-ray of $\text{H}_{2}$. According to \cite{2005ApJ...622..938G}, $k^{\prime}=1.2 \times 10^{-17} \mathrm{~cm}^{3} \mathrm{~s}^{-1}$ and $\zeta_{\mathrm{H}_{2}}=5.2 \times 10^{-17} \mathrm{~s}^{-1}$ are adopted. $n_{\mathrm{H} \mathrm{I}}$ and $n_{0}$ are the volume density of cold HI and total proton including atomic and molecular hydrogen, respectively. $\tau_{\mathrm{HI} \rightarrow \mathrm{H}_{2}}=\frac{1}{2 k^{\prime} n_{0}+\zeta_{\mathrm{H_{2}}}}$ is the timescale for cold HI to H$_2$ conversion. The fractional abundance of cold HI is defined as $x_{\mathrm{H} \mathrm{I}}=n_{\mathrm{H} \mathrm{I}}/n_{0}=n_{\mathrm{HI}}/(n_{\mathrm{HI}}+2n_{\mathrm{H}_2})$, by which the ages of different regions on and off the filaments can be estimated by Equation\,\ref{eq2}. $x_{\mathrm{H} \mathrm{I}}$ is calculated from [HI]/[H$_2$].

\par
In order to estimate the volume density of cold HI and $\text{H}_{2}$, we assume that the thickness is equal to the width for G28 and G47. Denoting the depth of filament as $L$,  the total proton density of atomic and molecular hydrogen can be calculated by $\bar{n}_{0}=\frac{N\left(\mathrm{H} \mathrm{I}\right)+2N\left(\mathrm{H}_{2}\right)}{L} $. The thickness of each filament is estimated by multiplying FAST beam $2'.9$ and the distance of the filament, which is the physical diameter of each circle region, to facilitate the comparison of ages between ON and OFF regions. The mean total proton density $\bar{n}_{0}$ of each region in G28 and G47 are listed in Table\,\ref{table1}.
\par
According to \cite{2005ApJ...622..938G}, when $k^{\prime} n_{0} \gg \zeta_{\mathrm{H_{2}}}$, $x_{\mathrm{HI}} \rightarrow \frac{\zeta_{\mathrm{H}_{2}}}{2 k^{\prime} n_{0}}$ and $n_{\mathrm{H}{\mathrm{I}}} \rightarrow \frac{\zeta_{\mathrm{H}_{2}}}{2 k^{\prime}}$ for a steady state. So the analytical model can only calculate the age of regions with $n_{\mathrm{H}{\mathrm{I}}}$ larger than $\frac{\zeta_{\mathrm{H}_{2}}}{2 k^{\prime}}$. The ages of 13 regions in G28 and 6 regions in G47 can be calculated (see Figure\,\ref{fig:chemical} and Table\,\ref{table1}). The ages of all 19 regions range from 2.2 to 13.2 Myr, similar to the timescale of 3.2 to 10.0 Myr for four dark interstellar clouds at 140 to 400\,pc \citep{2005ApJ...622..938G} and 5.0 to 10 Myr for five PGCCs \citep{2020RAA....20...77T}. G28 ON 3 and ON 4 are in the transition from unstable to steady state.

\par
Filaments have an uneven radial volume density distribution, which means the central regions are denser. So the above computation which assumes the volume density is evenly distributed along the line of sight will underestimate the volume density, and overestimate the filament ages. As a rough estimate, it is assumed that the radial distribution of filament volume density follows the 1D Plummer-like model\footnote{In this model, the radial volume density profile is approximately flat in the centre of the filament and drops as a power law in the envelope \citep[e.g.,][]{2008MNRAS.384..755N}.} \citep{1911MNRAS..71..460P}
\begin{equation}
n(r)=\frac{2 n_{0}}{1+\left(r / r_{0}\right)^{2}}, -L/2< r< L/2,
\end{equation}
in which $n_{0}$ is the central density and the fixed scale length $r_{0}$ is assumed to be $L/3$. In this case, $n_{0}\sim2.4\bar{n}_{0}$ and the ages of all 13 regions range from 0.9 to 4.5 Myr which is $\sim$30-40\% smaller than the original results. In \cite{2005ApJ...622..938G} and \cite{2020RAA....20...77T}, the cloud ages recovering central density are $\sim$50\% and $\sim$23\% smaller than original results, respectively.

\par
\begin{figure*}[htbp]
\epsscale{0.95}
\plotone{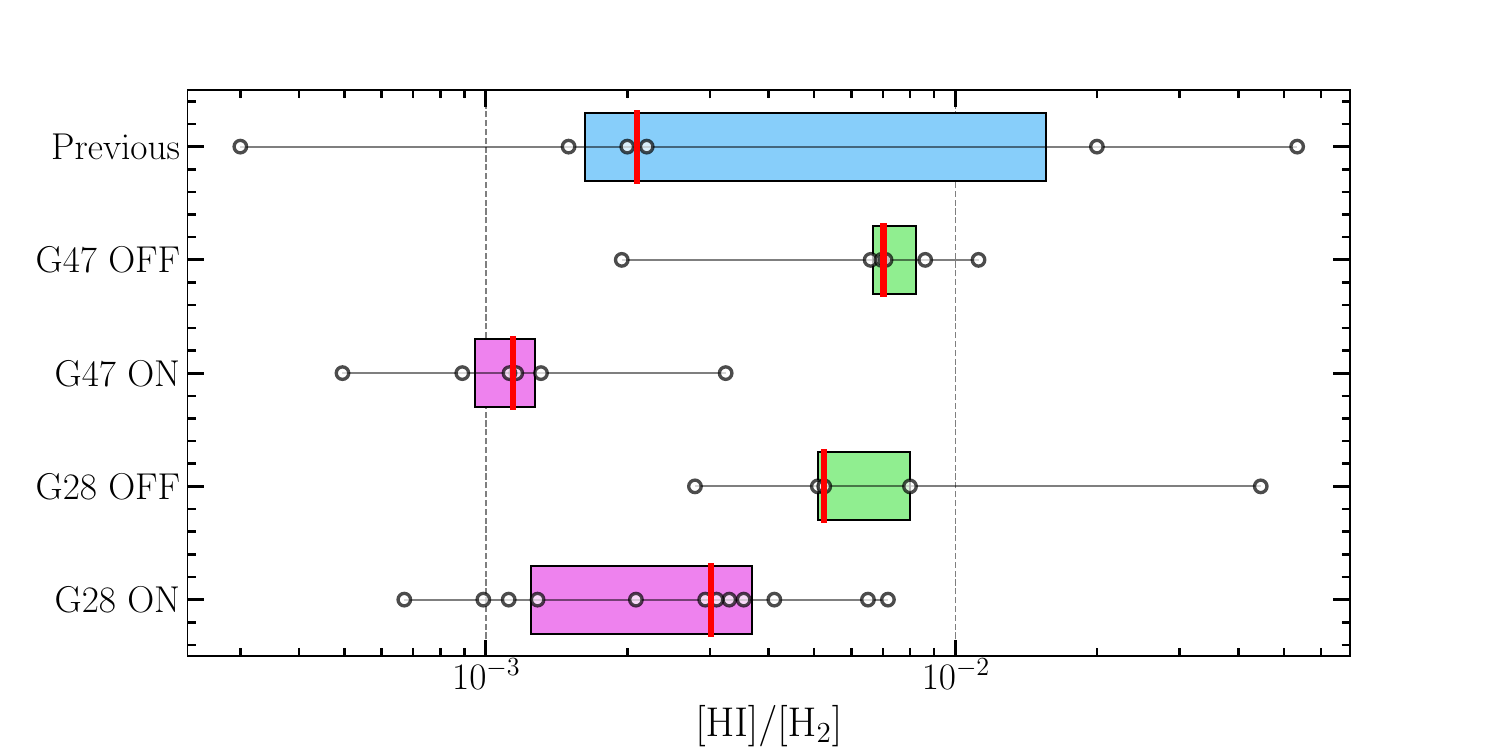}
\caption{[\hi]/[H$_2$] abundance ratios observed in G28 and G47 (including ON-filament and OFF-filament regions), as compared to previously reported values in the literature 
\citep[][mostly for nearby low-mass clouds]{2003ApJ...585..823L,2018ApJ...867...13Z,2020RAA....20...77T,2022A&A...658A.140L,2022ApJ...933L..26L}.
The horizontal lines connect minimal to maximum values, and each data points are marked as small open circles. The filled boxes encompass 25th percentile $Q_1$ to 75th percentile $Q_3$ of the corresponding [HI]/[H$_2$] data set, and the red vertical lines in the boxes mark median values.
}
\label{fig:HIbox}
\end{figure*}

From Table\,\ref{table1} and Figure\,\ref{fig:HIbox}, the values of [\hi]/[H$_{2}$] of OFF regions are generally higher than ON regions for both G28 and G47. This structure appears similar to the ``ring'' in \cite{2018ApJ...867...13Z} and smaller x$_{\mathrm{HI}}$ in the central beam than surrounding beams in \cite{2022ApJ...933L..26L}, implying the cold HI to H$_2$ conversion is further in central dense part of GMFs. The range of [HI]/[H$_{2}$] in this paper is similar to several previous results as plotted in Figure\,\ref{fig:HIbox}, which include nearby low-mass star forming molecular clouds (average [HI]/[H$_{2}$] of $1.5\times10^{-3}$ in \citealt{2003ApJ...585..823L}, [HI]/[H$_{2}$] between 0.2\% and 2\% in \citealt{2018ApJ...867...13Z}), PGCCs (average [HI]/[H] of $4.4\times10^{-3}$ in \citet{2020RAA....20...77T}, [HI]/[H$_{2}$] $\sim3\times10^{-4}$ varied by a factor $\sim3$ in \citealt{2022A&A...658A.140L}) and a relatively nearby ($d=1.8$ kpc) active high-mass star forming region G176.51+00.20 ([HI]/[H$_2$] from $\sim1.1\times10^{-3}$ to $\sim4.8\times10^{-3}$ in \citealt{2022ApJ...933L..26L}).

\begin{figure*}[htbp]
\gridline{\fig{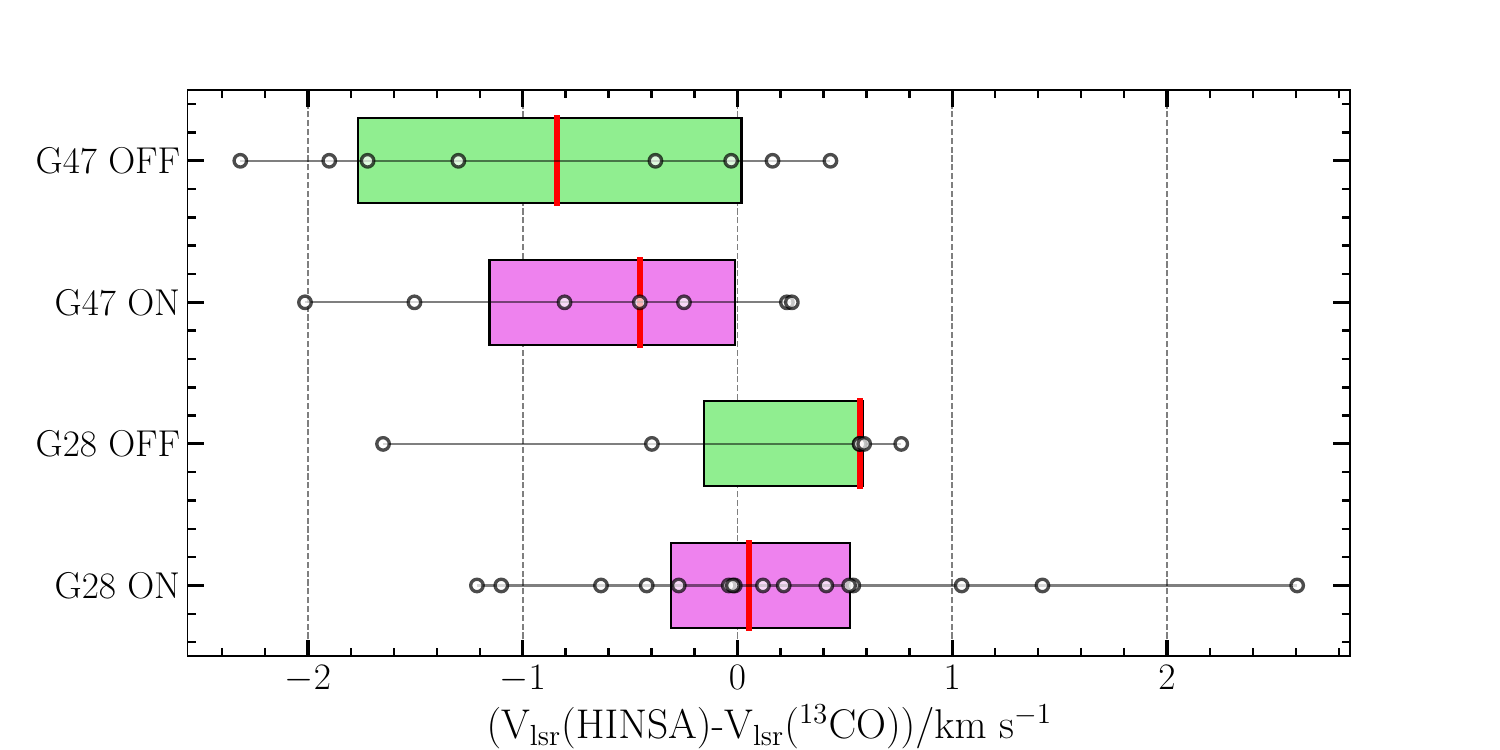}{0.52\textwidth}{(a)}
		  \fig{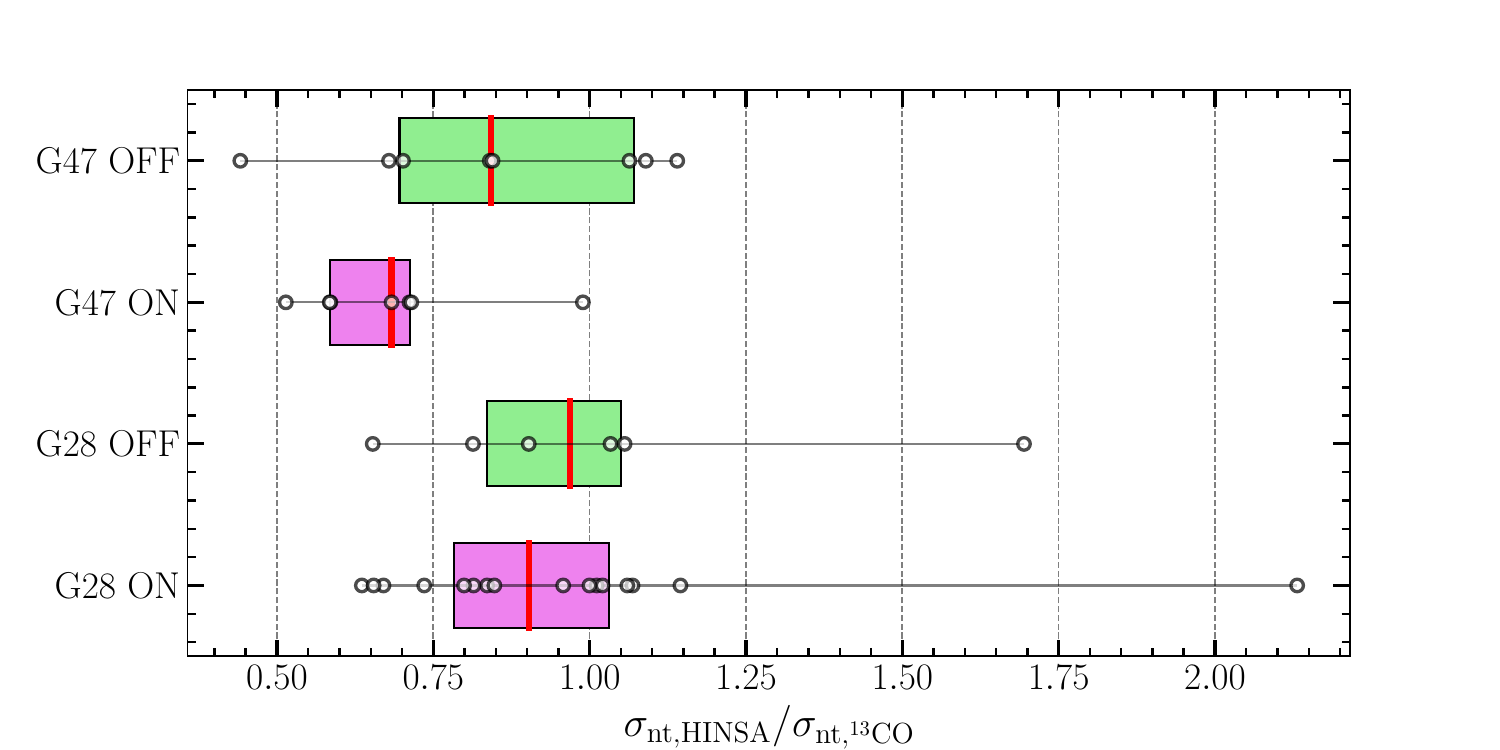}{0.52\textwidth}{(b)}
          }
\caption{Boxplot comparing the spectral line parameters of HINSA and $^{13}$CO of ON and OFF regions in G28 and G47. (a): Boxplot of the difference between the central velocities of HINSA and $^{13}$CO. (b): Boxplot of the ratio of the nonthermal velocity dispersion of HINSA to that of $^{13}$CO. The horizontal lines connect minimal to maximum values, and each data points of $V_{\mathrm{lsr}}(\mathrm{HINSA})-V_{\mathrm{lsr}}(^{13}\mathrm{CO})$ in panel (a) or $\sigma_{\mathrm{nt,HINSA}}-\sigma_{\mathrm{nt,^{13}CO}}$ in panel (b) are marked as small open circles. The filled boxes encompass 25th percentile $Q_1$ to 75th percentile $Q_3$ of the corresponding $V_{\mathrm{lsr}}(\mathrm{HINSA})-V_{\mathrm{lsr}}(^{13}\mathrm{CO})$ or $\sigma_{\mathrm{nt,HINSA}}-\sigma_{\mathrm{nt,^{13}CO}}$ data set, and the red vertical lines in the boxes mark median values.}
\label{fig:twobox}
\end{figure*}

\subsection{Comparable linewidth of HINSA and CO}
We quantitatively compare the central velocity and the nonthermal velocity dispersion of HINSA and $^{13}$CO. From Figure\,\ref{fig:twobox} (a), the difference of central velocity between HINSA and $^{13}$CO are mainly within $\pm$1 km s$^{-1}$. Figure\,\ref{fig:twobox} (b) shows the ratio of the nonthermal velocity dispersion of HINSA, $\sigma_{\mathrm{nt}, \mathrm{HINSA}}=\sqrt{\sigma_{\mathrm{HINSA}}^{2}-k T_{\mathrm{ex}} / m_{\mathrm{H}}}$, to that of $^{13}$CO, $\sigma_{\mathrm{nt},{ }^{13} \mathrm{CO}}=\sqrt{\sigma_{{ }^{13} \mathrm{CO}}^{2}-k T_{\mathrm{ex}} / m_{^{13} \mathrm{CO}}}$, where $k$ is the Boltzmann constant. $m_{\mathrm{H}}$ and $m_{^{13} \mathrm{CO}}$ are the mass of atomic hydrogen and $^{13}$CO molecule, respectively. The ratio
is slightly less than 1, which is consistent with the results in \cite{2003ApJ...585..823L} and \cite{2022ApJ...933L..26L}.  Both facts indicate that the cold HI traced by HINSA is mixed with the gas in cold, well-shielded regions of GMFs.

Most identified HINSA features show FWHM larger than 2\kms, while the linewidths of HINSA in nearby molecular clouds are $\lesssim 1 \mathrm{~km} \mathrm{~s}^{-1}$ \citep{2003ApJ...585..823L,2020RAA....20...77T,2022A&A...658A.140L}. The wide lines of the HINSA features are likely due to the large distance of the bones from us. A FAST beam covers a much larger physical scale in distant clouds than nearby ones, leading to larger velocity dispersion. According to \cite{Larson1981}, physical size $R$ is scaled with velocity dispersion as $\sigma \sim R^{0.33}$. For typical GMCs at a factor of 10 times larger distance than nearby clouds, the linewidth is a factor of 2.13 times larger, consistent with the HINSA linewidth we measured in distant GMCs.


\subsection{Simultaneous cloud formation and star formation}
Our study reveals ongoing atomic-to-molecular conversion in the two bones at timescales of few to ten Myr. On the other hand, there are ongoing massive star formation in the two clouds \citep{2015MNRAS.450.4043W,2018A&A...609A..43X,2019RAA....19..183X}. It is therefore of great interest to compare the timescales of cloud formation and star formation.

We estimate free-fall timescale of each circular region as
\begin{equation}
t_{\mathrm{ff}}=\left(\frac{3 \pi}{32 G \rho}\right)^{\frac{1}{2}},
\end{equation}
in which $\rho$ is the mass volume density. Considering $\rho \approx \rho\left ( \mathrm{HINSA} \right ) +\rho\left ( \mathrm{H}_{2} \right )$, we derive that $t_{\mathrm{ff}}$ of ON regions in G28 and G47 is $\sim (0.9$ to $2.5)$ Myr and $t_{\mathrm{ff}}$ of OFF regions in G28 and G47 is $\sim (1.2$ to $2.7)$ Myr. We also estimate the ``cross-filament'' timescale $t_{\mathrm{cr}}$ calculated as $L/c_{\mathrm{s}}$. The acoustic speed $c_{\mathrm{s}}$ is 
\begin{equation}
c_{\mathrm{s}}=\sqrt{\frac{k T}{\mu m_{\mathrm{p}}}},
\end{equation}
in which $T$ is the temperature of filament estimated as $T_{\mathrm{dust}}$ in Table\,\ref{table1}, $\mu=2.37$ for filament\footnote{$\mu=2.37$ is the mean molecular weight for per ``free particle'' for a molecular cloud made with $N(\mathrm{H})/N(\mathrm{He})=10$ and negligible metals \citep{WangKe2014}.} and $m_{\mathrm{p}}$ is proton mass. $t_{\mathrm{cr}}$ is $\sim (14.2$ to $16.3)$ Myr for ON regions in G28 and G47 and $\sim (14.1$ to $15.6)$ Myr for OFF regions.

The above timescales for massive star formation are comparable to the timescales for cloud formation, suggesting that we are witnessing a rapid star formation process launched almost simultaneously as the clouds are forming. In fact, G28 hosts an IRDC G28.53-0.25 with early signs of star formation \citep{2015ApJ...805..171L}.
A similar conclusion has been reached by the HISA study towards the relatively nearby and evolved (compared to G28 and G47) DR1 filament in Cygnus X \citep{LiC2023_CygX_HI}.
This ``simultaneous cloud formation and star formation'' scenario is consistent with the statistical study by \citet{Urquhart22_EvoTrend}, where $\sim$10,000 ATLASGAL clumps are classified into four evolutionary stages. By counting the relative number population of different evolutionary stages, they find that the prestellar stage is relatively short compared to protostellar stages, suggesting that star formation has been launched very early.
Moreover, the scenario is also consistent with SMA and ALMA studies of the earliest stages, in which researchers attempted to identify the initial conditions right before star formation, but all so far have turned out to find star formation activities with deep enough observations \citep[e.g.,][]{WangKe2011,WangKe2014,qz15,Feng2016,Sanhueza2017,Sanhueza2019_ASHES,Pillai2019,Jiao_2023}. These results again suggest an early start of star formation and a highly dynamic process and mass assembly of massive protocluster formation \citep{assemble,quarks2}.

The ``fast star formation'' scenario, if further verified, would challenge massive star formation models that requires previously formed low-mass stars to feedback the clouds and elevate equivalent Jeans mass \citep[e.g.,][]{2002Natur.416...59M,2005Natur.438..332K,2007ApJ...656..959K}, because low-mass stars form in longer timescales.
We also note that the neighborhood of these clouds show HII regions, Class I, II, and III YSOs and stellar clusters \citep{2019RAA....19..183X}. The relation of these features and the forming molecular clouds G28 and G47 need to be studied further. In the case of DR 21 filament, for example, expanding HII regions in Cygnus-X compressed and shaped an atomic cloud to form DR 21, which itself is actively forming stars \citep{LiC2023_CygX_HI}. In that case, the HII regions are from an older generation of star formation, while DR 21 hosts the current generation of star formation. A similar star formation history could happen for G28 and G47. Besides, the molecular forming process naturally result in a gas inflow, which could explain the filamentary inflows revealed by recent ALMA observations \citep[e.g.][]{peretto13,atoms15}. Their connections should serve as the bridge between GMFs and local star formation.  Encouraged by the pioneer case study of G28 and G47, we are observing more GMFs with a wider range of physical parameters, and will investigate this new scenario further in a statistically significant sample. 

\section{Summary}\label{sec:summary}
\par
Using sensitive FAST HI mapping observations, we have found ubiquitous HINSA features in two distant GMFs. It is the first time HINSA is observed towards bones of the Milky Way. The main results are as follows.
\begin{enumerate}
\item[1.] We extracted 38 HINSA components from 29 regions, among which 21 components have a line width $>3$km s$^{-1}$, comparable or slightly narrower than the CO line width. The characteristic trace cold conversion of HI to H$_2$ conversion in the two GMFs, revealing an ubiquity of HINSA towards distant GMCs and Galactic ``bones'' that have a larger line width than the canonical 1\kms~line width of HINSA in nearby clouds.
The column density of HINSA of 18 ON regions ranges from $\sim$(0.4 to 14.1)$\times10^{19}$ cm$^{-2}$. $N$(HINSA) of 11 OFF regions ranges from $\sim$(1.8 to 32.8)$\times10^{19}$ cm$^{-2}$. The values of [HI]/[H$_2$] of OFF regions ($\sim$(1.9 to 44.7)$\times10^{-3}$) are generally higher than ON regions ($\sim$(0.5 to 7.2)$\times10^{-3}$), implying further atomic-to-molecular transition in central dense part of GMFs.

\item[2.] Using an analytical model, we derived the chemical ages (i.e., starting from the first net HI-H$_2$ conversion) of the two clouds to be in the range of 2.2 to 13.2 Myr. The cloud formation timescales are comparable to the observed star formation timescales in these two GMFs, suggesting a an almost simultaneous star formation and cloud formation. Massive stars are forming without previous generation of low-mass stars, challenging some of the leading star formation models.

\end{enumerate}

With the ubiquity of HINSA (and not HISA) in GMCs and Galctic ``bones'',
this study demonstrates FAST's huge potential in cloud formation, and will inspire more observations towards GMFs to further investigate cloud formation in connection with star formation. Furthermore, SOFIA has observed ordered magnetic field lines perpendicular to the spine of the G47 \citep{2022ApJ...926L...6S}, suggesting that B-fields likely play a role in supporting the G47 from collapse. The extraordinary HINSA features ($>15$ K) identified in this study are excellent for Zeeman splitting observations to directly measure magnetic field strength much more efficiently than traditional ways \citep{2022Natur.601...49C}, which we are carrying out with FAST.

\begin{acknowledgments}
We thank the referee for valuable and constructive suggestions that helped improve the original manuscript.
We thank the FAST staff for their support during on-site observations and data management.
We acknowledge support from the National Science Foundation of China (12041305, 12033005), 
the Tianchi Talent Program of Xinjiang Uygur Autonomous Region, the China-Chile Joint Research Fund (CCJRF No. 2211),
and the High-Performance Computing Platform of Peking University.
CCJRF is provided by Chinese Academy of Sciences South America Center for Astronomy (CASSACA) and established by National Astronomical Observatories, Chinese Academy of Sciences (NAOC) and Chilean Astronomy Society (SOCHIAS) to support China-Chile collaborations in astronomy.
Sun was supported by the Beijing Training Program of Innovation for the Undergraduate Student Research Study at Peking University.
This publication makes use of molecular line data from the Boston University-FCRAO Galactic Ring Survey (GRS). The GRS is a joint project of Boston University and Five College Radio Astronomy Observatory, funded by the National Science Foundation under grants AST-9800334, AST-0098562, \& AST-0100793. This publication makes use of data from FUGIN, FOREST Unbiased Galactic plane Imaging survey with the Nobeyama 45-m telescope, a legacy project in the Nobeyama 45-m radio telescope.

\end{acknowledgments}

\setlength{\tabcolsep}{1.5pt}
\begin{longrotatetable}
\begin{deluxetable*}{cccccccccccccccc}
\renewcommand{\arraystretch}{1.2}
\tablecaption{Derived parameters of HINSA features in G28 and G47}
\tablewidth{700pt}
\tabletypesize{\scriptsize}
\tablehead{
\colhead{Position} & \colhead{RA(J2000)} & \colhead{Dec(J2000)} & \colhead{Comp} & \colhead{$\Delta V(^{13}\mathrm{CO)}$} & \colhead{V$_{\mathrm{LSR}}(^{13}\mathrm{CO)}$}& \colhead{$\tau_{0}$} & \colhead{$\Delta V$} & \colhead{V$_{\mathrm{LSR}}$} & \colhead{$T_{\mathrm{ab}}$} & \colhead{$T_{\mathrm{dust}}$} & \colhead{$N\left(\mathrm{HINSA}\right)$} & \colhead{$N\left(\mathrm{H}_{2}\right)$} & \colhead{[HI]/[H$_{2}$]} & \colhead{$\bar{n}_{0}$} & \colhead{Timescale}  \\ 
\colhead{} & \colhead{} & \colhead{} & \colhead{} & \colhead{(km s$^{-1}$)} & 
\colhead{(km s$^{-1}$)} & \colhead{} & \colhead{(km s$^{-1}$)} & \colhead{(km s$^{-1}$)} & \colhead{(K)} & \colhead{(K)} &\colhead{(10$^{19}$ cm$^{-2}$)} & \colhead{(10$^{21}$ cm$^{-2}$)} & \colhead{(10$^{-3}$)} & \colhead{($10^3$ cm$^{-3}$)} & \colhead{(Myr)}
} 
\decimalcolnumbers
\startdata
		G28-ON 1 & 281.22 & -4.03 & 1 & 3.43$\pm$0.04 & 90.73$\pm$0.02&0.24$\pm$0.07 & 2.94$\pm$0.35 & 90.72$\pm$0.18 & 12.03$\pm$0.008 &21.5 & 2.1$\pm$0.5 & 16.2 & 1.3 & 2.55 & --\\
		G28-ON 2 & 281.16 & -4.03 & 1 &3.76$\pm$0.10 & 88.56$\pm$0.04& 0.60$\pm$0.23 & 8.03$\pm$1.32 & 87.46$\pm$0.80 & 42.15$\pm$0.050 &19.4 & 14.1$\pm$4.6 & 19.6 & 7.2 & 3.09 & 2.5$\pm$0.3 \\
		G28-ON 3 & 281.09 & -4.03 & 1 &4.01$\pm$0.07 & 87.87$\pm$0.03& 0.24$\pm$0.07 & 4.11$\pm$0.50 & 87.82$\pm$0.25 & 15.73$\pm$0.010 & 18.2 & 2.9$\pm$0.7 & 25.8 & 1.1 & 4.06 & close to steady state\\
		G28-ON 4 & 281.07 & -3.99 & 1 & 4.96$\pm$0.05 & 86.35$\pm$0.02 &0.24$\pm$0.09 & 4.10$\pm$0.65 & 87.39$\pm$0.33 & 14.95$\pm$0.009 & 17.6 & 2.8$\pm$0.9 & 41.9 & 0.7 & 6.59 & close to steady state \\
		G28-ON 5 & 281.12 & -3.96 & 1 & 3.37$\pm$0.04 & 88.64$\pm$0.01 &0.21$\pm$0.07 & 3.50$\pm$0.49 & 88.37$\pm$0.25 & 14.75$\pm$0.008 &19.4 & 2.1$\pm$0.6 & 21.6 & 1.0 & 3.40 & -- \\
		\multirow{2}{*}{G28-ON 6} & \multirow{2}{*}{281.17} & \multirow{2}{*}{-3.91} & 1 & 3.01$\pm$0.03 & 88.55$\pm$0.01 &0.29$\pm$0.06 & 3.28$\pm$0.25 &88.12$\pm$0.14 & 19.96$\pm$0.019  & \multirow{2}{*}{19.4} & 2.8$\pm$0.5 & \multirow{2}{*}{19.5} & \multirow{2}{*}{3.3} & \multirow{2}{*}{3.07} & \multirow{2}{*}{3.0$\pm$0.3} \\ & & & 2 & 3.51$\pm$0.12 & 94.64$\pm$0.05 & 0.36$\pm$0.08 &3.43$\pm$0.26 &95.06$\pm$0.15 & 21.13$\pm$0.018 & &3.6$\pm$0.6 &\\
		\multirow{2}{*}{G28-ON 7} & \multirow{2}{*}{281.21} & \multirow{2}{*}{-3.87} & 1 & 3.00$\pm$0.07 & 87.78$\pm$0.03 & 0.19$\pm$0.04 &2.63$\pm$0.22 &88.32$\pm$0.11 & 13.92$\pm$0.023  & \multirow{2}{*}{19.8} & 1.5$\pm$0.3 & \multirow{2}{*}{17.2} & \multirow{2}{*}{4.1} & \multirow{2}{*}{2.71}  & \multirow{2}{*}{3.3$\pm$0.4}\\ & & &2 & 6.72$\pm$0.25 & 94.29$\pm$0.10 &0.42$\pm$0.10 &4.55$\pm$0.44 &94.81$\pm$0.25 & 25.52$\pm$0.018 & &5.6$\pm$1.1 &\\
		\multirow{2}{*}{G28-ON 8} & \multirow{2}{*}{281.22} & \multirow{2}{*}{-3.82} & 1 & 3.52$\pm$0.06 & 87.19$\pm$0.03 & 0.15$\pm$0.03 &2.34$\pm$0.21 &88.61$\pm$0.10 & 10.15$\pm$0.051 &   \multirow{2}{*}{20.7} & 1.0$\pm$0.2 & \multirow{2}{*}{26.1} & \multirow{2}{*}{3.1} & \multirow{2}{*}{4.11} & \multirow{2}{*}{2.2$\pm$0.2}  \\ & & &2 & 5.69$\pm$0.14 & 95.52$\pm$0.06 &0.64$\pm$0.09 &3.78$\pm$0.18 &95.64$\pm$0.12 & 34.14$\pm$0.038 & &7.1$\pm$0.8 &\\
		\multirow{2}{*}{G28-ON 9} & \multirow{2}{*}{281.22} & \multirow{2}{*}{-3.77} & 1 & 3.52$\pm$0.07 & 87.08$\pm$0.03 &0.27$\pm$0.09 &3.79$\pm$0.49 &89.68$\pm$0.24 & 17.92$\pm$0.056 & \multirow{2}{*}{20.1} & 3.0$\pm$0.8 & \multirow{2}{*}{18.6} & \multirow{2}{*}{6.5} & \multirow{2}{*}{2.93} & \multirow{2}{*}{2.7$\pm$0.3}\\ & & &2& 4.89$\pm$0.07 & 95.18$\pm$0.03 &0.85$\pm$0.11 &3.66$\pm$0.16 &95.40$\pm$0.12 & 41.56$\pm$0.054 & &9.1$\pm$1.0 &\\
		G28-ON 10 & 281.23 & -3.71 & 1 &3.94$\pm$0.04 & 95.51$\pm$0.02 & 0.40$\pm$0.07 &3.22$\pm$0.21 &95.49$\pm$0.12 & 21.04$\pm$0.015 &  20.3 & 3.7$\pm$0.6 & 17.8 & 2.1 & 2.80 & 3.9$\pm$0.5 \\
		G28-ON 11 & 281.24 & -3.54 & 1 & 3.77$\pm$0.05 & 95.54$\pm$0.02 &0.47$\pm$0.09 &3.83$\pm$0.27 &94.90$\pm$0.17 & 24.99$\pm$0.020 & 20.3 & 5.2$\pm$0.9 & 14.7 & 1.9 & 4.0$\pm$0.5 \\
		G28-ON 12 & 281.24 & -3.54 & 1 & 4.89$\pm$0.06 & 96.07$\pm$0.02&0.24$\pm$0.15 & 5.63$\pm$1.61 & 94.57$\pm$0.79 & 15.24$\pm$0.009 & 20.3 & 3.9$\pm$2.1 & 13.3 & 2.9 & 2.09 & 4.9$\pm$1.6 \\
		G28-OFF 1 & 281.41 & -3.97 & 1 & 8.71$\pm$0.28 & 84.44$\pm$0.12& 1.58$\pm$0.34 & 7.12$\pm$0.61 & 85.01$\pm$0.44 & 44.48$\pm$0.145 & 19.7 & 32.8$\pm$6.0 & 7.3 & 44.7 & 1.18 & 4.4$\pm$0.9\\
		G28-OFF 2 & 281.35 & -3.95 & 1 & 3.88$\pm$0.17 & 86.85$\pm$0.07& 0.51$\pm$0.17 & 4.07$\pm$0.51 & 87.61$\pm$0.32 & 15.44$\pm$0.021 & 19.6 & 6.0$\pm$1.7 & 7.5 & 8.0 & 1.19 & 6.8$\pm$1.7 \\
		\multirow{2}{*}{G28-OFF 3} & \multirow{2}{*}{281.27} & \multirow{2}{*}{-3.93} & 1 & 7.14$\pm$0.38 & 86.63$\pm$0.13 &0.17$\pm$0.10 &4.71$\pm$2.59 &84.98$\pm$0.63 & 10.94$\pm$0.010 & \multirow{2}{*}{19.9} & 2.4$\pm$1.5 & \multirow{2}{*}{10.4} & \multirow{2}{*}{5.1} & \multirow{2}{*}{1.64} & \multirow{2}{*}{5.4$\pm$1.1}\\ & & &2& 2.88$\pm$0.12 & 95.66$\pm$0.05  & 0.32$\pm$0.08 &3.11$\pm$0.29 &95.26$\pm$0.16 & 16.65$\pm$0.012 &  &2.9$\pm$0.6 &\\
		G28-OFF 4 & 281.04 & -3.94 & 1 & 4.48$\pm$0.07 & 86.78$\pm$0.03 &0.52$\pm$0.09 & 4.10$\pm$0.30 & 87.35$\pm$0.18 & 30.19$\pm$0.046 &19.1 & 6.2$\pm$0.9 & 22.3 & 2.8 & 3.51 & 2.7$\pm$0.2\\
		G28-OFF 5 & 281.00 & -3.94 & 1 & 2.47$\pm$0.08 & 86.47$\pm$0.03 &0.75$\pm$0.14 & 4.23$\pm$0.33 & 87.06$\pm$0.21 & 30.61$\pm$0.064 & 19.9 & 9.3$\pm$1.5 & 17.7 & 5.3 & 2.79 & 3.0$\pm$0.3 \\
		G47-ON 1 & 289.18 & 12.92 & 1 & 3.32$\pm$0.10 & 56.96$\pm$0.04& 0.05$\pm$0.02 & 2.45$\pm$0.47 & 56.71$\pm$0.21 & 4.87$\pm$0.001 & 18.5 & 0.4$\pm$0.1 & 7.1 & 0.5 & 1.23 & -- \\
		G47-ON 2 & 289.19 & 12.81 & 1 & 5.29$\pm$0.12 & 58.18$\pm$0.05 &0.18$\pm$0.07 & 5.28$\pm$0.95 & 57.72$\pm$0.44 & 18.21$\pm$0.010 &19.0 & 3.0$\pm$1.1 & 9.3 & 3.2 & 1.61 & 6.7$\pm$2.4 \\
				\multirow{2}{*}{G47-ON 3} & \multirow{2}{*}{289.18} & \multirow{2}{*}{12.69} & 1 & 3.95$\pm$0.14 & 56.20$\pm$0.05 & 0.06$\pm$0.03 & 2.78$\pm$0.57 & 55.39$\pm$0.25 & 5.74$\pm$0.001 & \multirow{2}{*}{18.4} & 0.5$\pm$0.2 & \multirow{2}{*}{8.6} & \multirow{2}{*}{0.9} & \multirow{2}{*}{1.49} & \multirow{2}{*}{--} \\
		&  & & 2 & 2.93$\pm$0.37 & 62.03$\pm$0.15 & 0.03$\pm$0.02 & 2.20$\pm$0.69 & 60.52$\pm$0.30 & 2.65$\pm$0.001 & & 0.2$\pm$0.1 & \\
		G47-ON 4 & 289.14 & 12.58 & 1 & 4.92$\pm$0.14 & 58.72$\pm$0.06 &0.08$\pm$0.03 & 2.62$\pm$0.40 & 58.95$\pm$0.18 & 7.59$\pm$0.002 &18.9 & 0.7$\pm$0.2 & 6.0 & 1.2 & 1.04  & --\\
		G47-ON 5 & 289.02 & 12.52 & 1 & 4.81$\pm$0.08 & 58.37$\pm$0.03 & 0.08$\pm$0.03 & 2.90$\pm$0.49 & 58.62$\pm$0.22 & 7.36$\pm$0.002 & 18.6 & 0.7$\pm$0.3 & 5.7 & 1.3 & 0.98 & --\\
		G47-ON 6 & 288.97 & 12.53 & 1 &4.44$\pm$0.09 & 56.72$\pm$0.04 &  0.06$\pm$0.03 & 2.68$\pm$0.58 & 54.70$\pm$0.26 & 5.47$\pm$0.001 & 18.7 &0.6$\pm$0.3 &  4.9 & 1.1 & 0.85 & --\\
		\multirow{2}{*}{G47-OFF 1} & \multirow{2}{*}{289.19} & \multirow{2}{*}{12.58} & 1 & 3.05$\pm$0.09 & 57.16$\pm$0.04 & 0.17$\pm$0.04 & 2.65$\pm$0.28 & 55.86$\pm$0.12 & 14.42$\pm$0.005 & \multirow{2}{*}{19.3} & 1.4$\pm$0.3 & \multirow{2}{*}{9.4} & \multirow{2}{*}{1.9} & \multirow{2}{*}{1.63}& \multirow{2}{*}{--}\\
		&  & & 2 & 2.97$\pm$0.24 & 61.30$\pm$0.10 & 0.06$\pm$0.03 & 2.19$\pm$0.45 & 59.40$\pm$0.18 & 5.52$\pm$0.006 & & 0.4$\pm$0.2\\
		\multirow{2}{*}{G47-OFF 2} & \multirow{2}{*}{289.23} & \multirow{2}{*}{12.58} & 1 & 4.66$\pm$0.55 & 56.40$\pm$0.22 &0.25$\pm$0.05 & 3.23$\pm$0.24 & 56.02$\pm$0.12 & 21.52$\pm$0.012 &\multirow{2}{*}{19.2} & 2.6$\pm$0.5 &  \multirow{2}{*}{4.9} & \multirow{2}{*}{6.6} & \multirow{2}{*}{0.86}& \multirow{2}{*}{11.0$\pm$4.0}\\
		&  & & 2 & 2.06$\pm$0.37 & 61.75$\pm$0.16 & 0.09$\pm$0.03 & 2.34$\pm$0.30 & 59.43$\pm$0.12 & 7.91$\pm$0.016 & &0.6$\pm$0.2& \\
		G47-OFF 3 & 289.28 & 12.58 & 1 & 5.17$\pm$1.11 & 58.65$\pm$0.47 & 0.24$\pm$0.09 & 5.54$\pm$0.89 & 56.93$\pm$0.44 & 21.30$\pm$0.012 & 18.9 &4.3$\pm$1.5 & 5.0 & 8.6 & 0.88 & 9.5$\pm$3.4 \\
		G47-OFF 4 & 289.22 & 12.53 & 1 & 3.56$\pm$0.24 & 55.94$\pm$0.09 &0.32$\pm$0.04 & 3.08$\pm$0.16 & 56.10$\pm$0.09 & 24.52$\pm$0.019 & 18.9 & 3.2$\pm$0.4 & 4.6 & 7.0 & 0.80 & 11.9$\pm$4.6\\
		\multirow{2}{*}{G47-OFF 5} & \multirow{2}{*}{289.24} & \multirow{2}{*}{12.49} & 1 & 4.07$\pm$0.75 & 55.97$\pm$0.29 & 0.20$\pm$0.07 & 4.69$\pm$0.73 & 56.40$\pm$0.36 & 15.09$\pm$0.006 & \multirow{2}{*}{18.8} & 2.9$\pm$1.0 & \multirow{2}{*}{4.2} & \multirow{2}{*}{7.1}  & \multirow{2}{*}{0.74}& \multirow{2}{*}{13.2$\pm$6.6}\\
		&  & & 2 & 2.19$\pm$0.22 & 60.62$\pm$0.09 & 0.02$\pm$0.01 & 1.18$\pm$0.25 & 60.59$\pm$0.11 & 1.42$\pm$0.013 & & 0.1$\pm$0.0 &\\
		\multirow{2}{*}{G47-OFF 6} & \multirow{2}{*}{289.06} & \multirow{2}{*}{12.47} & 1 & \multirow{2}{*}{6.07$\pm$0.48} & \multirow{2}{*}{58.03$\pm$0.20 }& 0.22$\pm$0.06 & 3.38$\pm$0.38 & 54.78$\pm$0.17 & 17.15$\pm$0.012 & \multirow{2}{*}{18.9} &2.3$\pm$0.6 & \multirow{2}{*}{4.1} & \multirow{2}{*}{11.2} & \multirow{2}{*}{0.71}& \multirow{2}{*}{11.0$\pm$4.5}\\
		&  & & 2 & & & 0.19$\pm$0.06 & 3.61$\pm$0.49 & 58.60$\pm$0.21 & 15.76$\pm$0.010 & & 2.3$\pm$0.7 &\\
\enddata
\end{deluxetable*}
\label{table1}
\end{longrotatetable}

\appendix
\label{Appendix}
\FloatBarrier

\begin{figure*}
\gridline{\fig{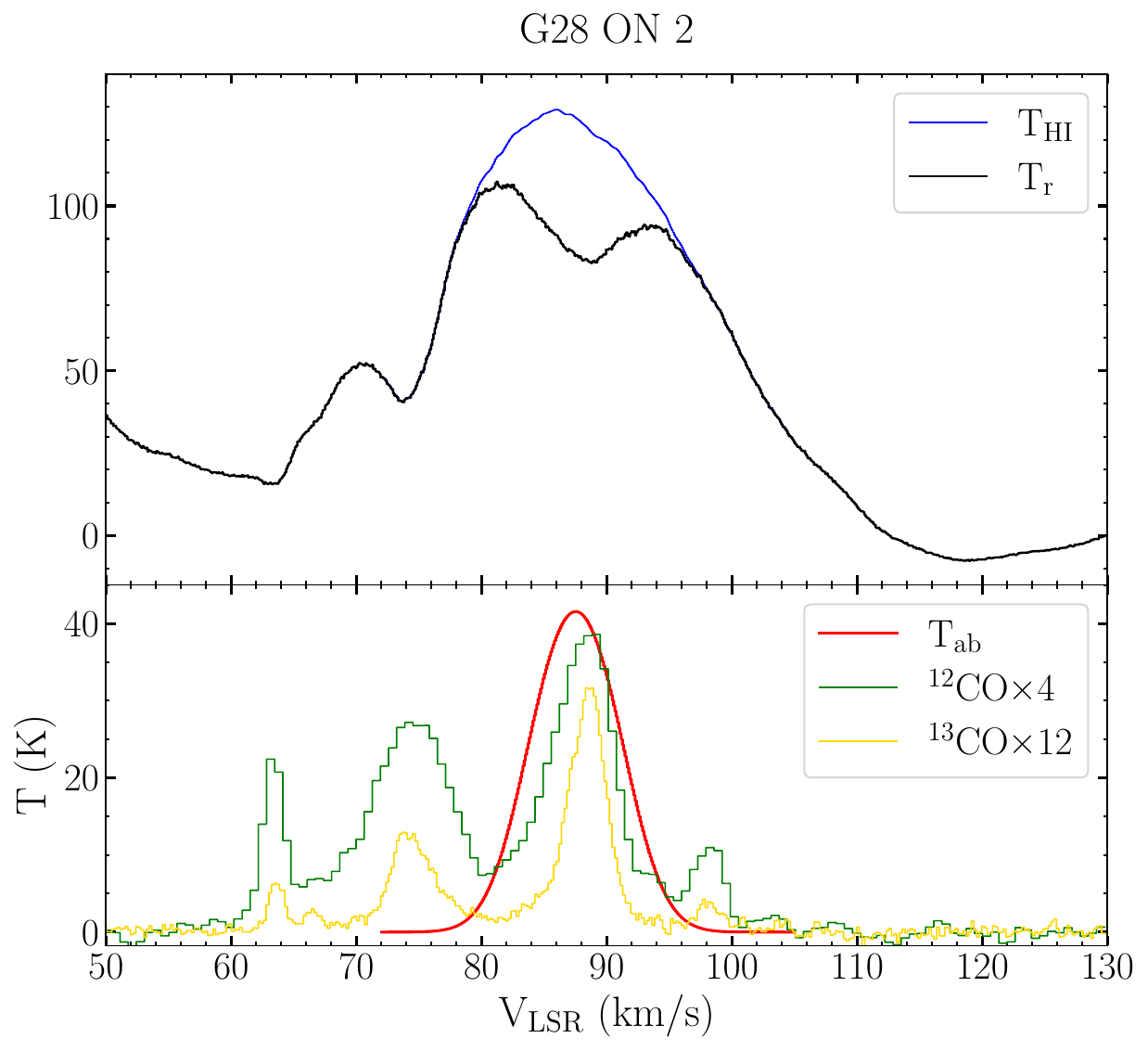}{0.33\textwidth}{(a) ON 2}
          \fig{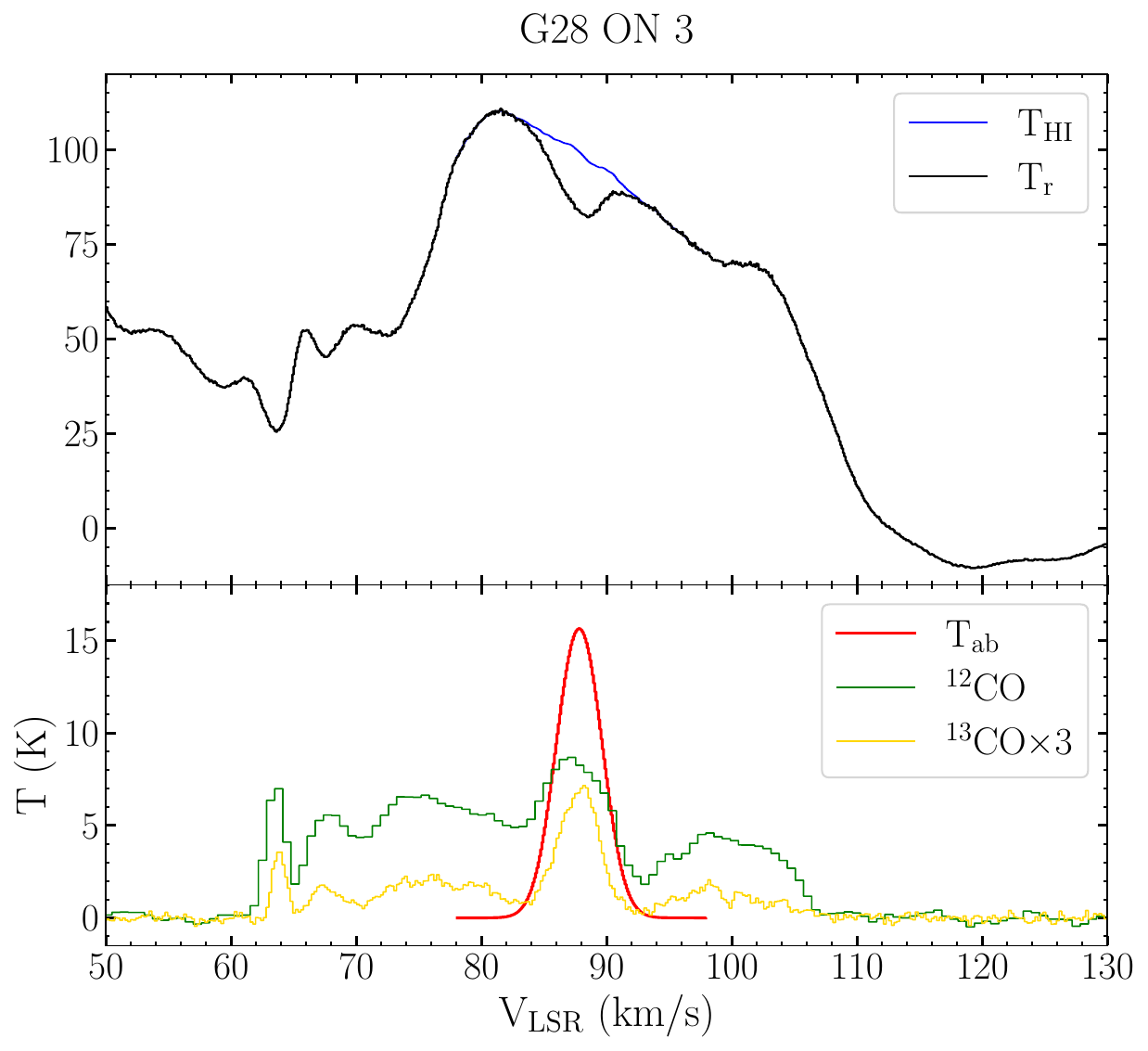}{0.33\textwidth}{(b) ON 3}
          \fig{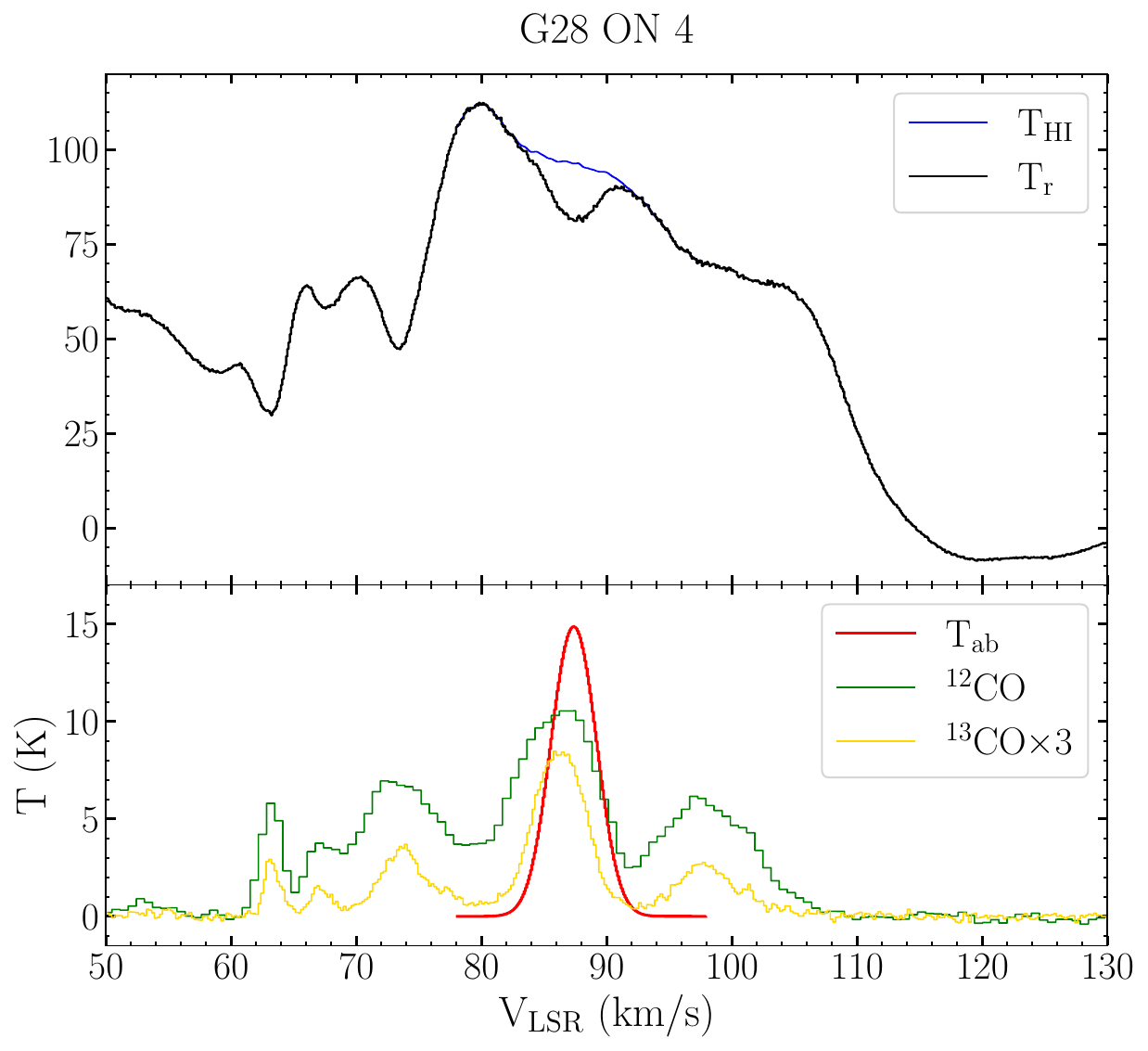}{0.33\textwidth}{(c) ON 4}
          }
\gridline{
		  \fig{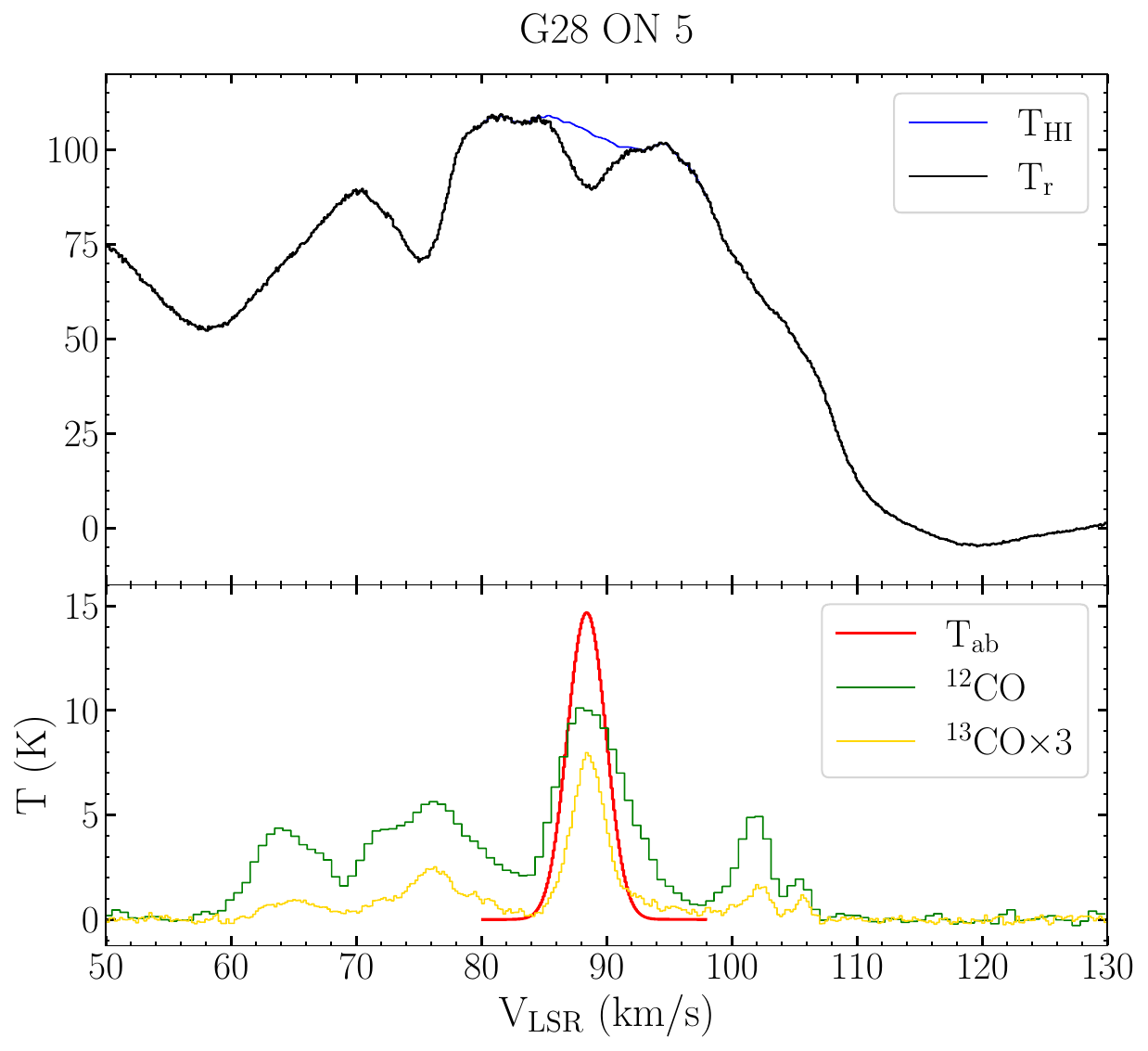}{0.33\textwidth}{(d) ON 5}
		  \fig{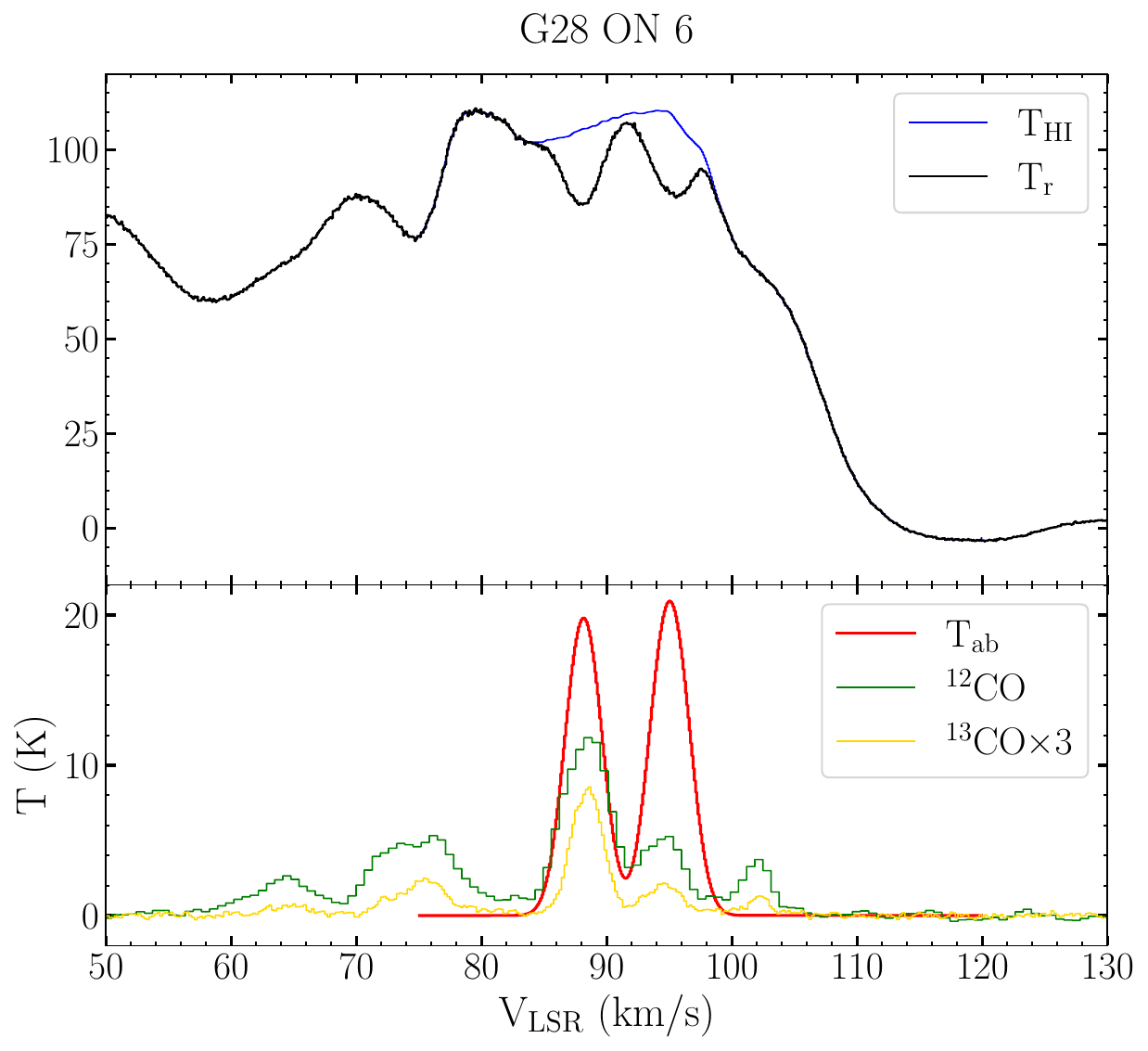}{0.33\textwidth}{(e) ON 6}
		  \fig{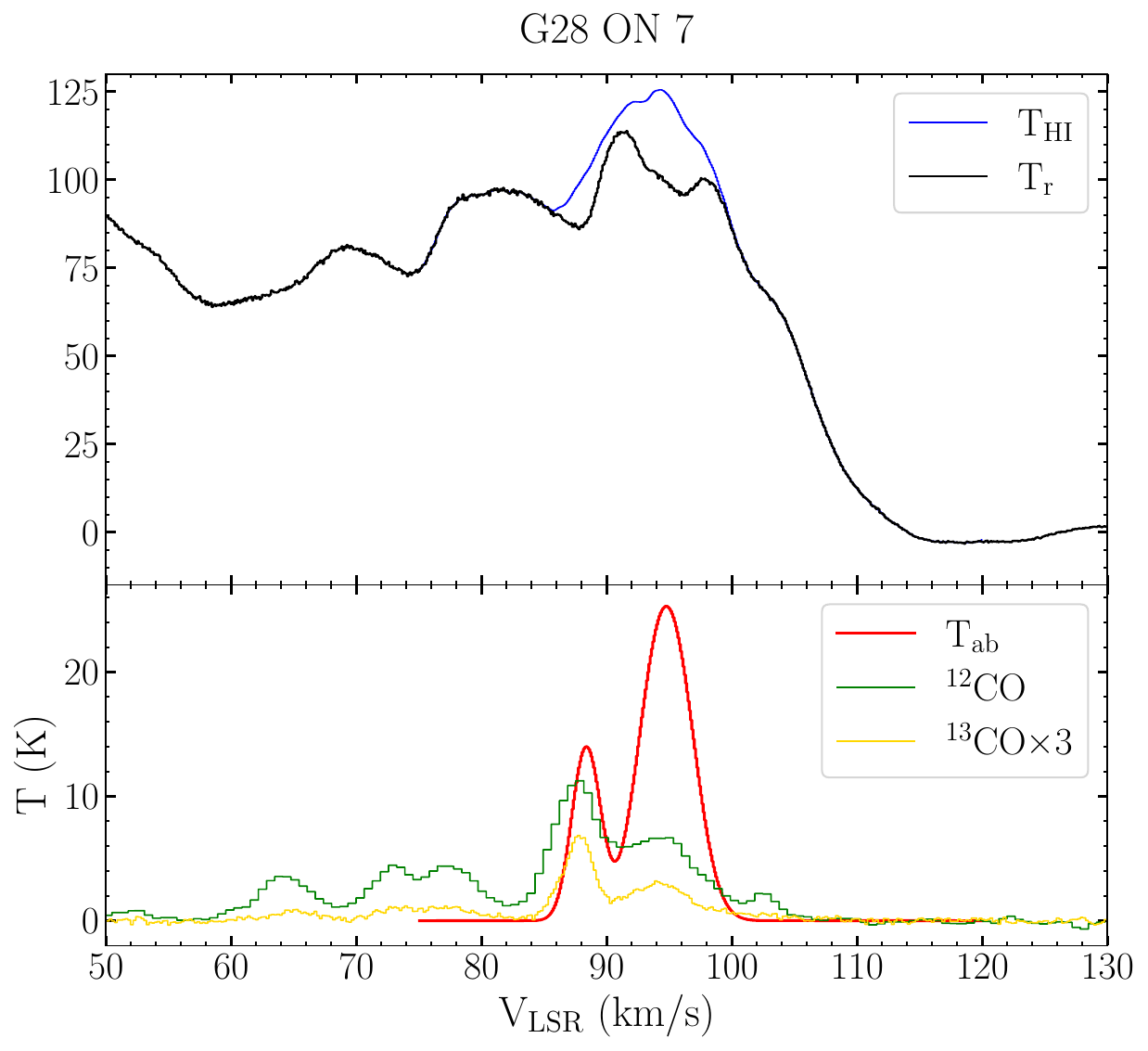}{0.33\textwidth}{(f) ON 7}
		  }
\gridline{
		  \fig{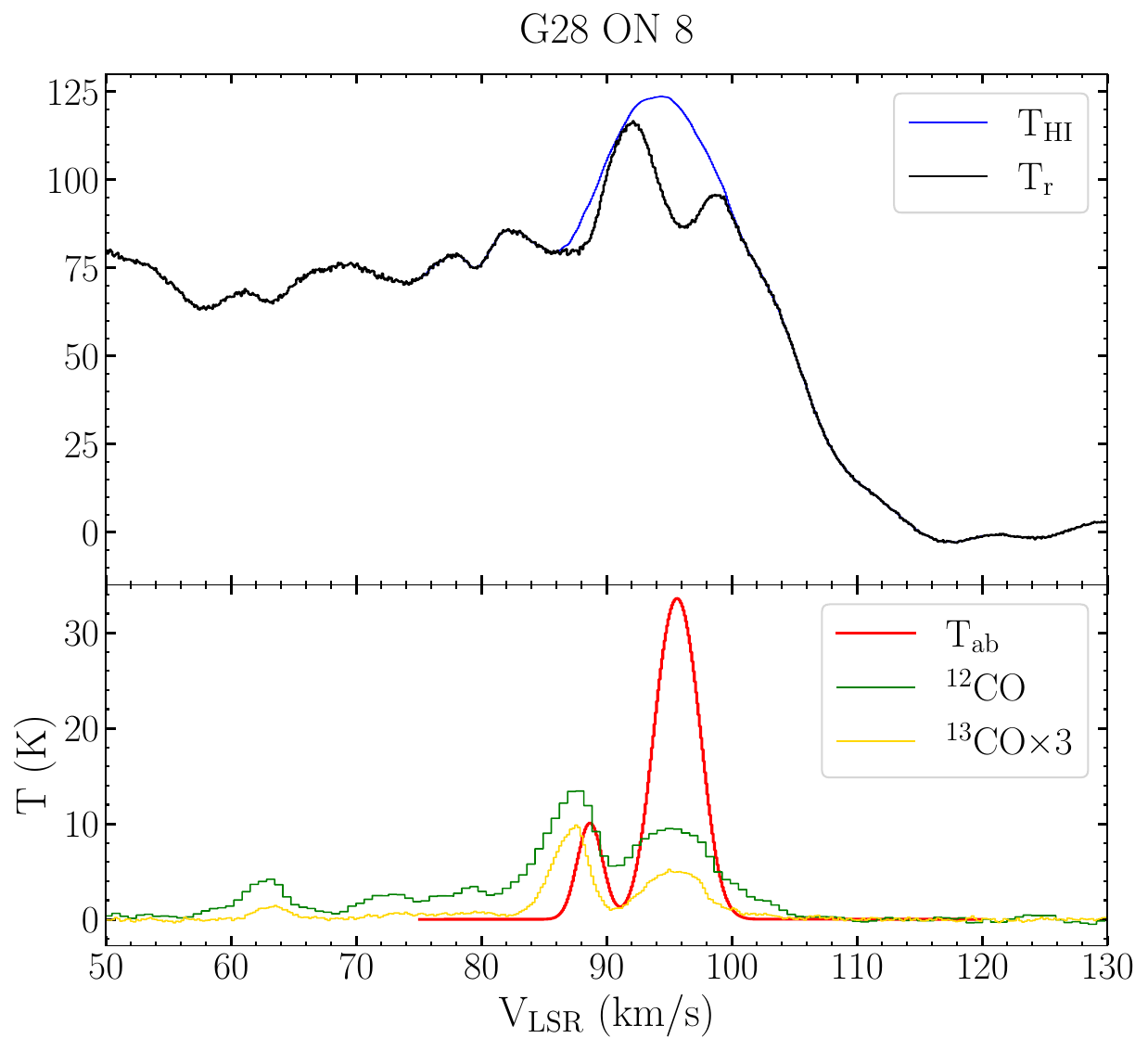}{0.33\textwidth}{(g) ON 8}
		  \fig{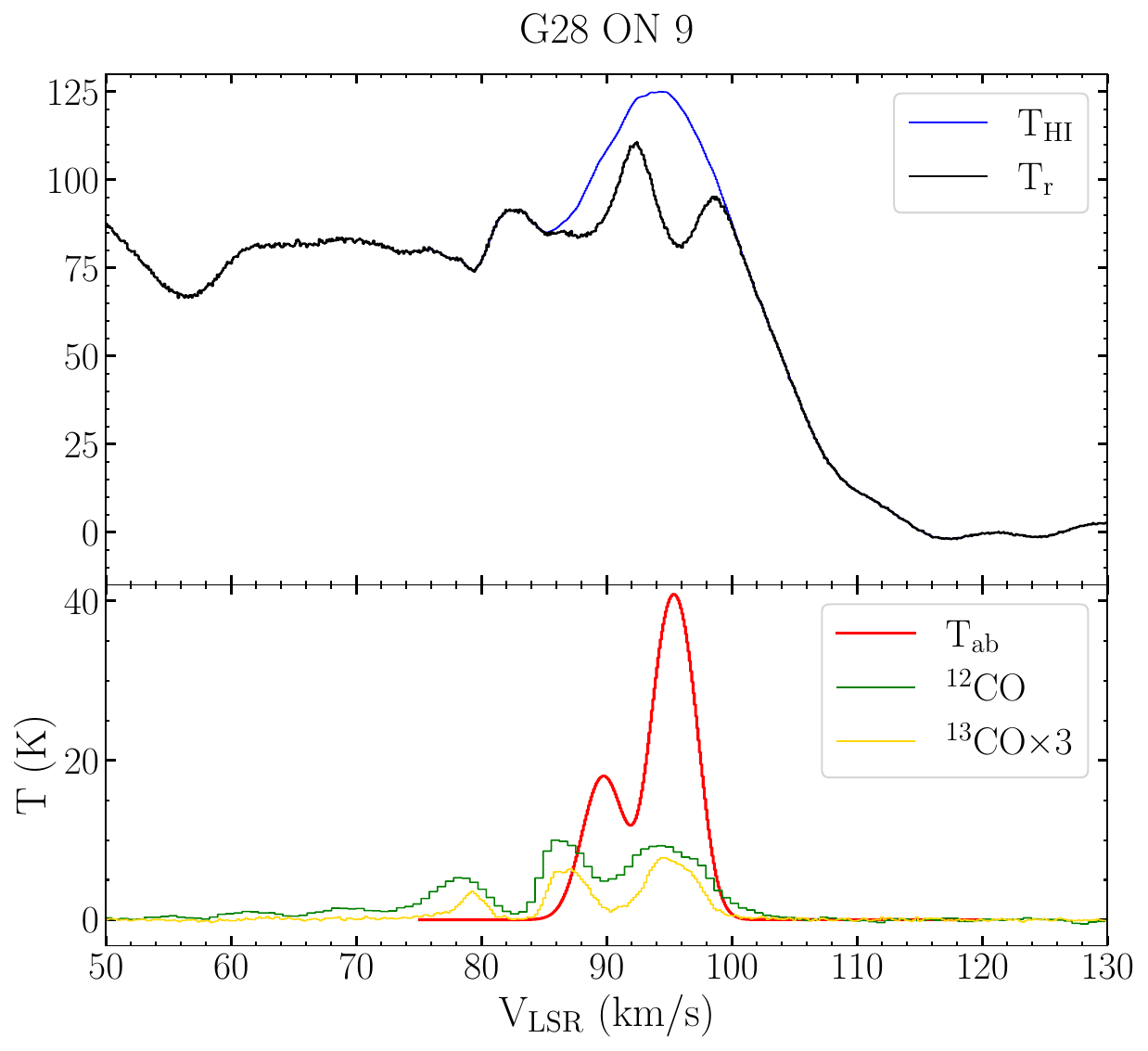}{0.33\textwidth}{(h) ON 9}
		  \fig{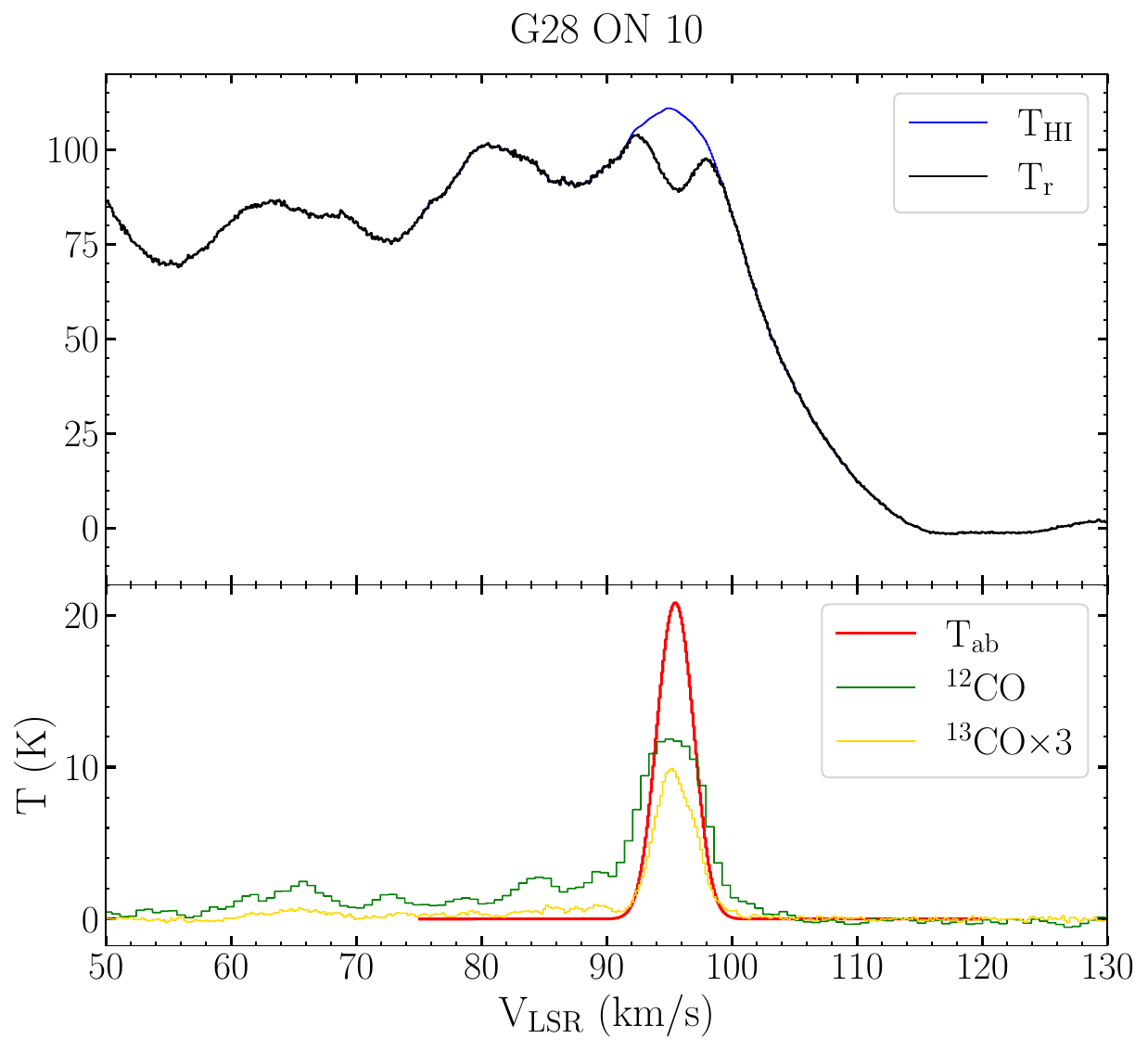}{0.33\textwidth}{(i) ON 10}
		  }
\end{figure*}
\begin{figure*}
\gridline{
		  \fig{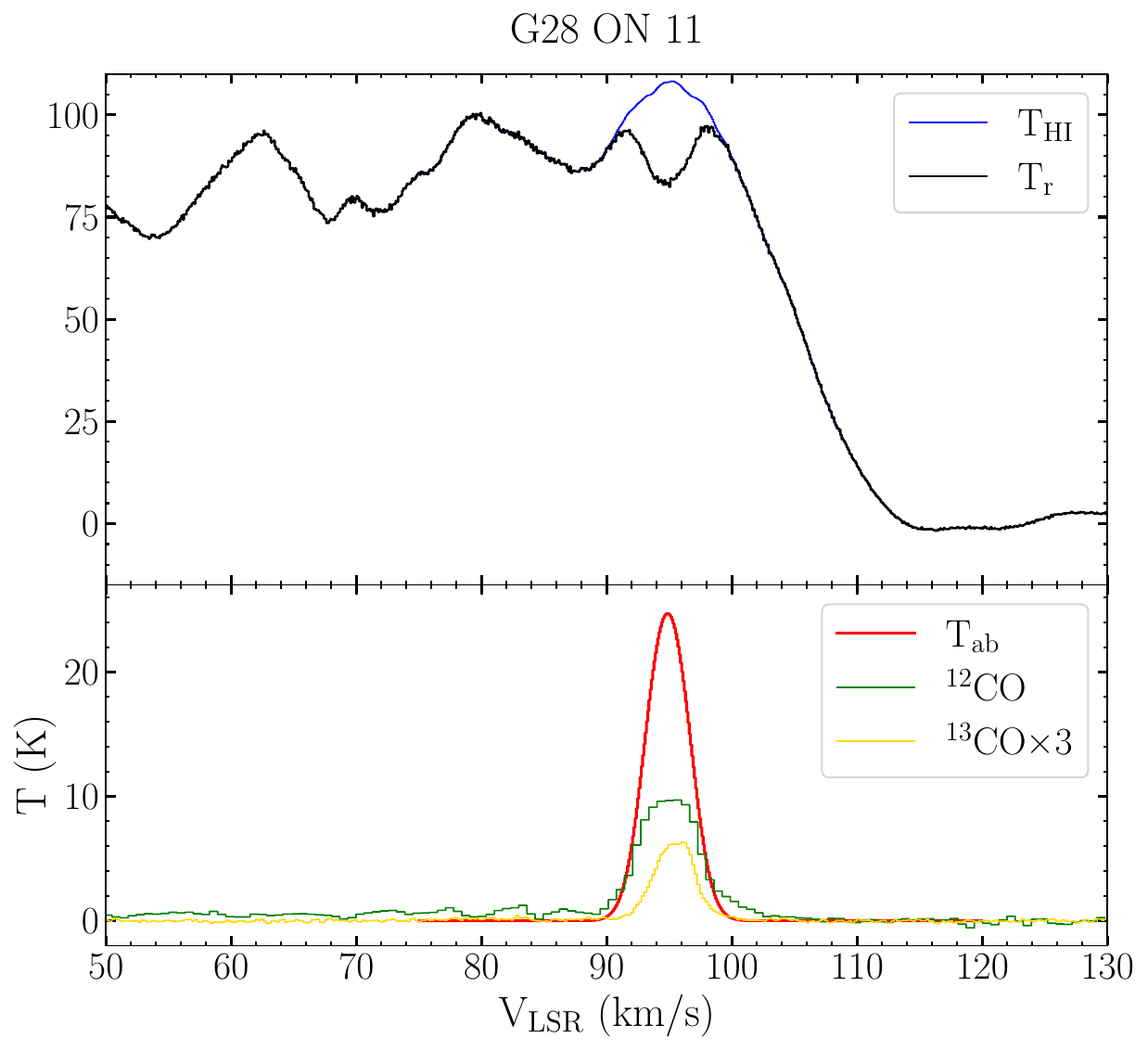}{0.33\textwidth}{(j) ON 11}
		  \fig{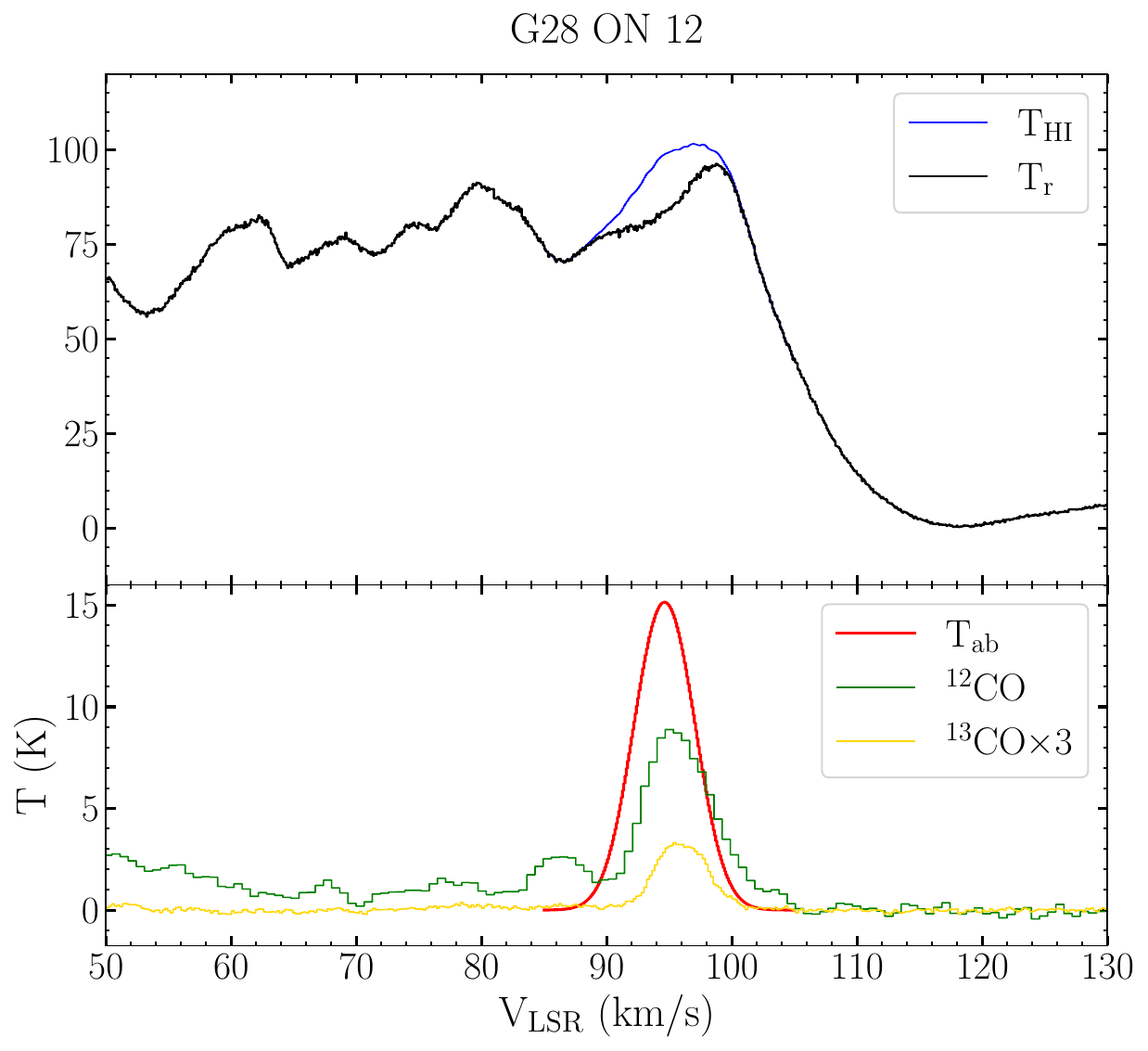}{0.33\textwidth}{(k) ON 12}
		  \fig{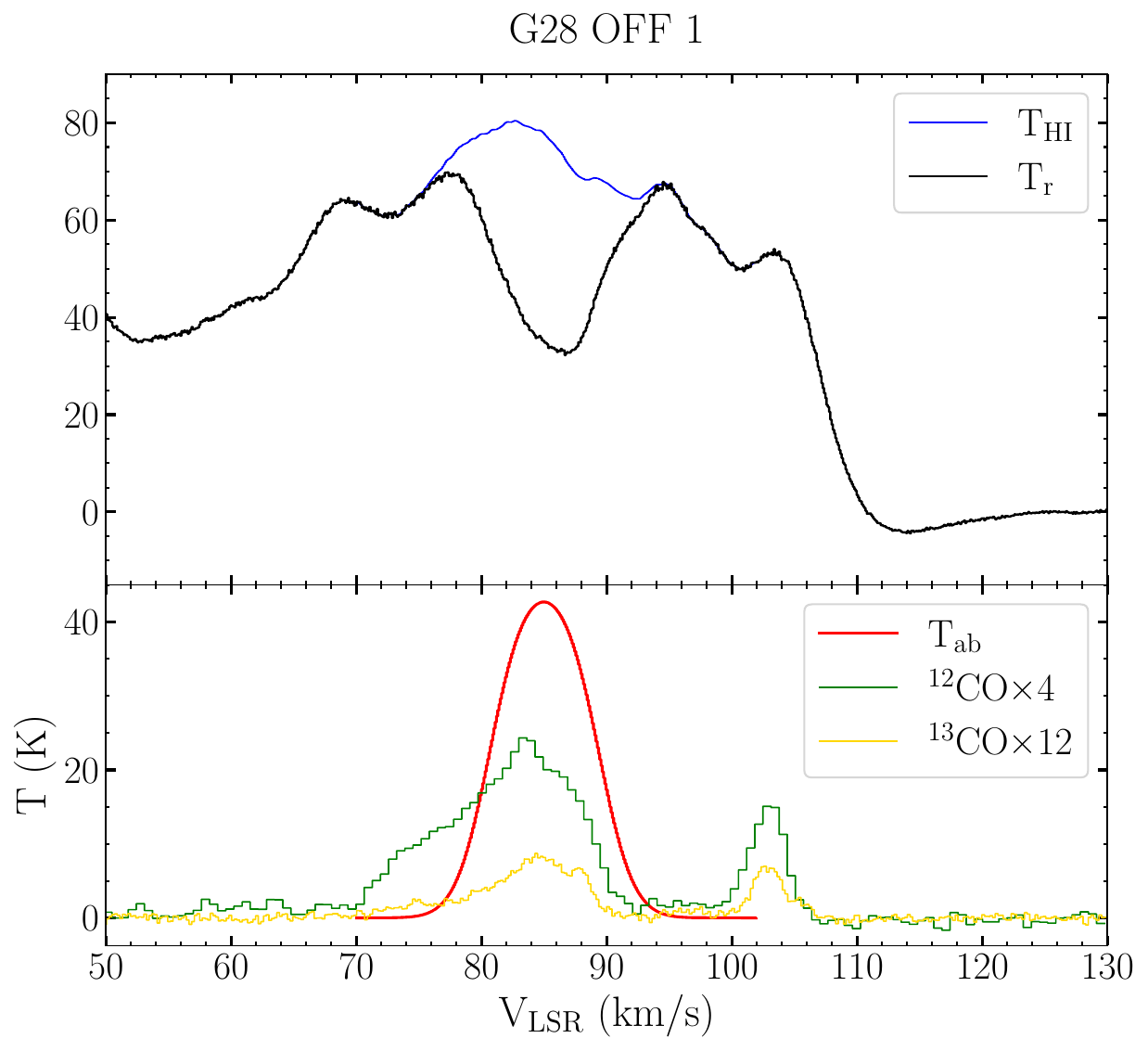}{0.33\textwidth}{(l) OFF 1}
		  }
\end{figure*}
\begin{figure*}
\gridline{
		  \fig{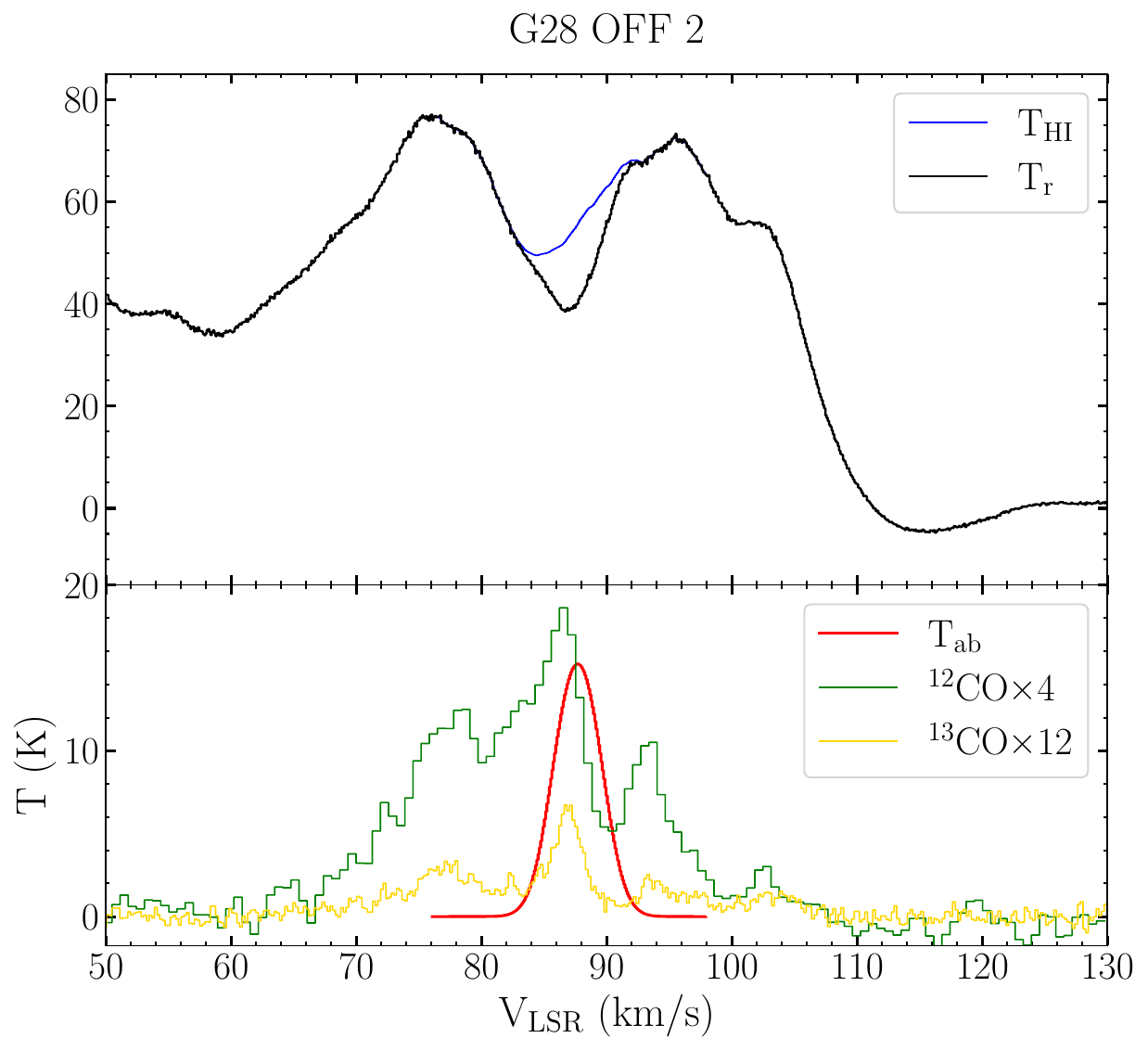}{0.33\textwidth}{(m) OFF 2}
		  \fig{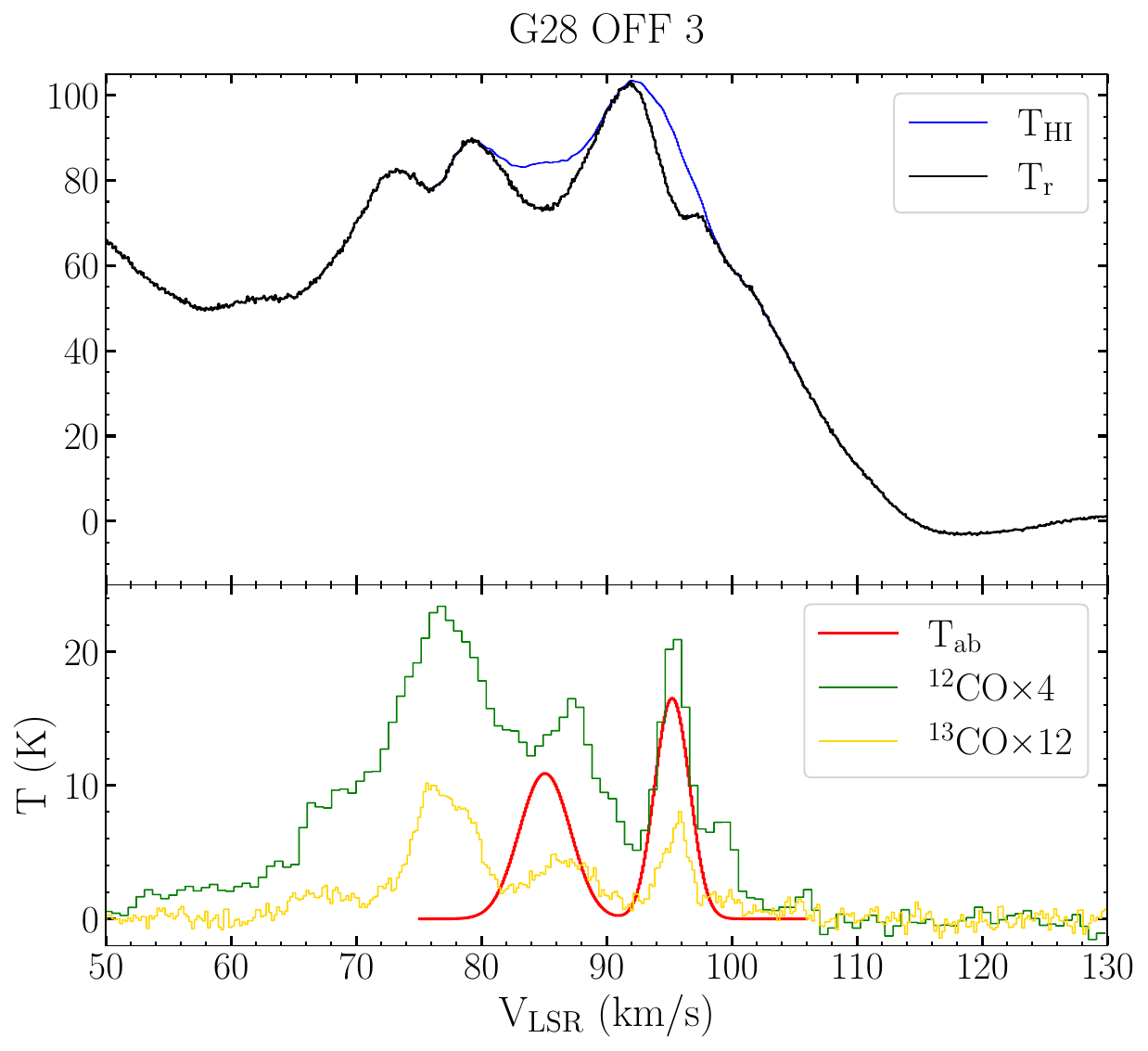}{0.33\textwidth}{(n) OFF 3}
		  \fig{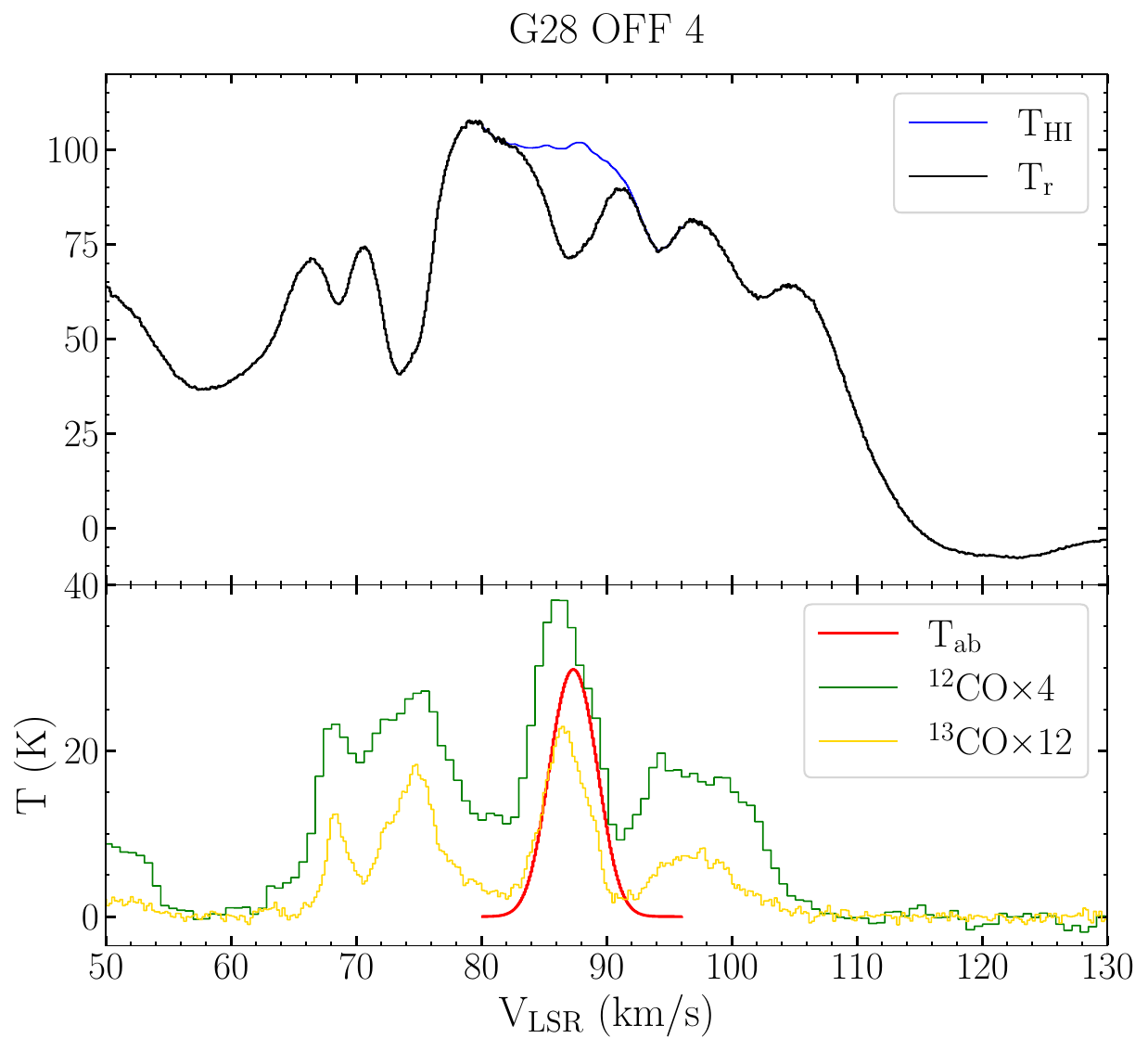}{0.33\textwidth}{(o) OFF 4}
		  }

\gridline{
		  \fig{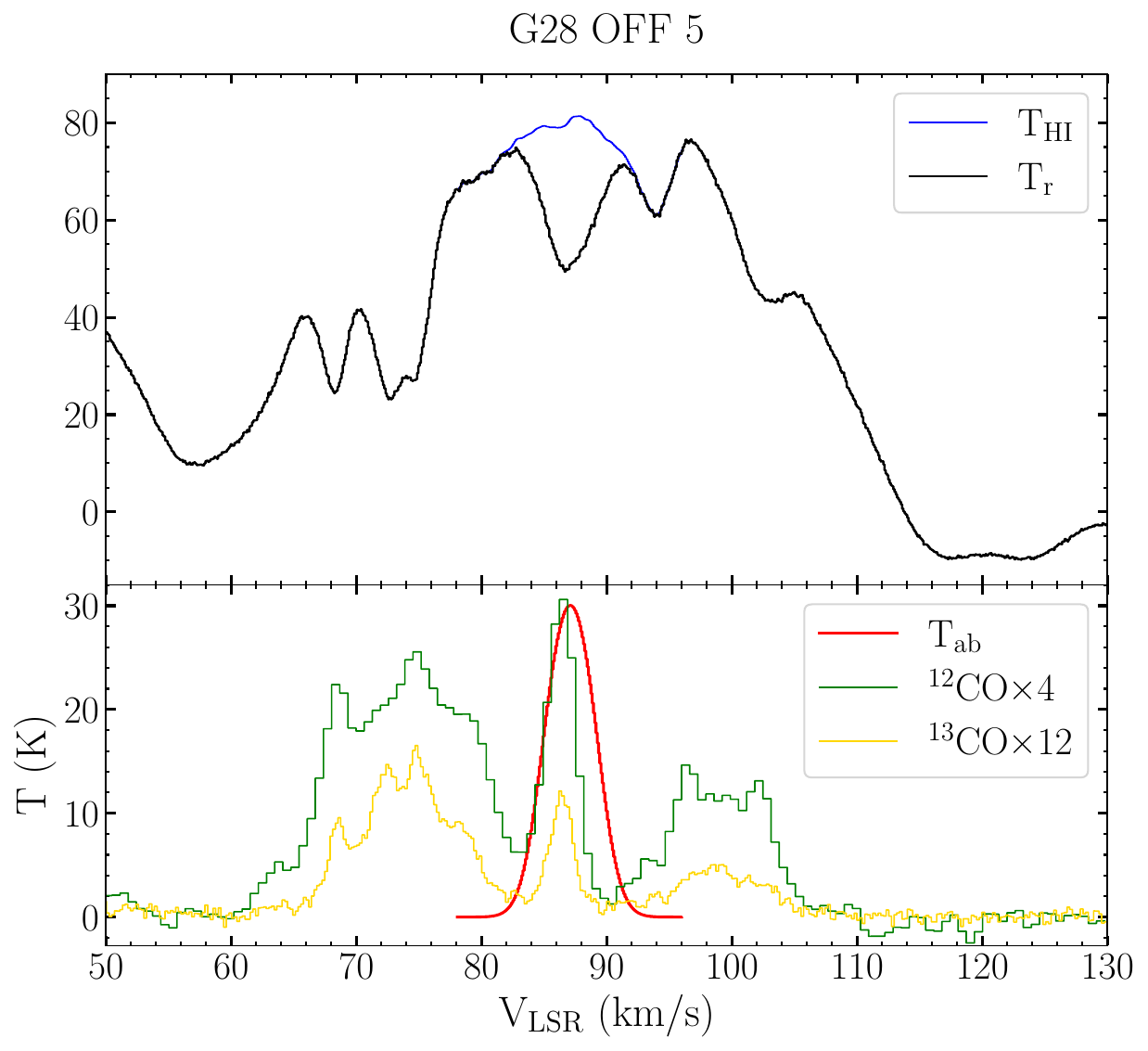}{0.33\textwidth}{(p) OFF 5}
		  }
\caption{HINSA fitting and CO spectra towards G28. Descriptions are the same as Figure\,\ref{fig:4Figs}. For ON regions, the temperature of $^{13}$CO is magnified 3 times. For OFF regions, the temperatures of $^{12}$CO and $^{13}$CO are magnified 4 and 12 times, respectively.}
\label{fig:HINSA G28}
\end{figure*}

\begin{figure*}
\gridline{\fig{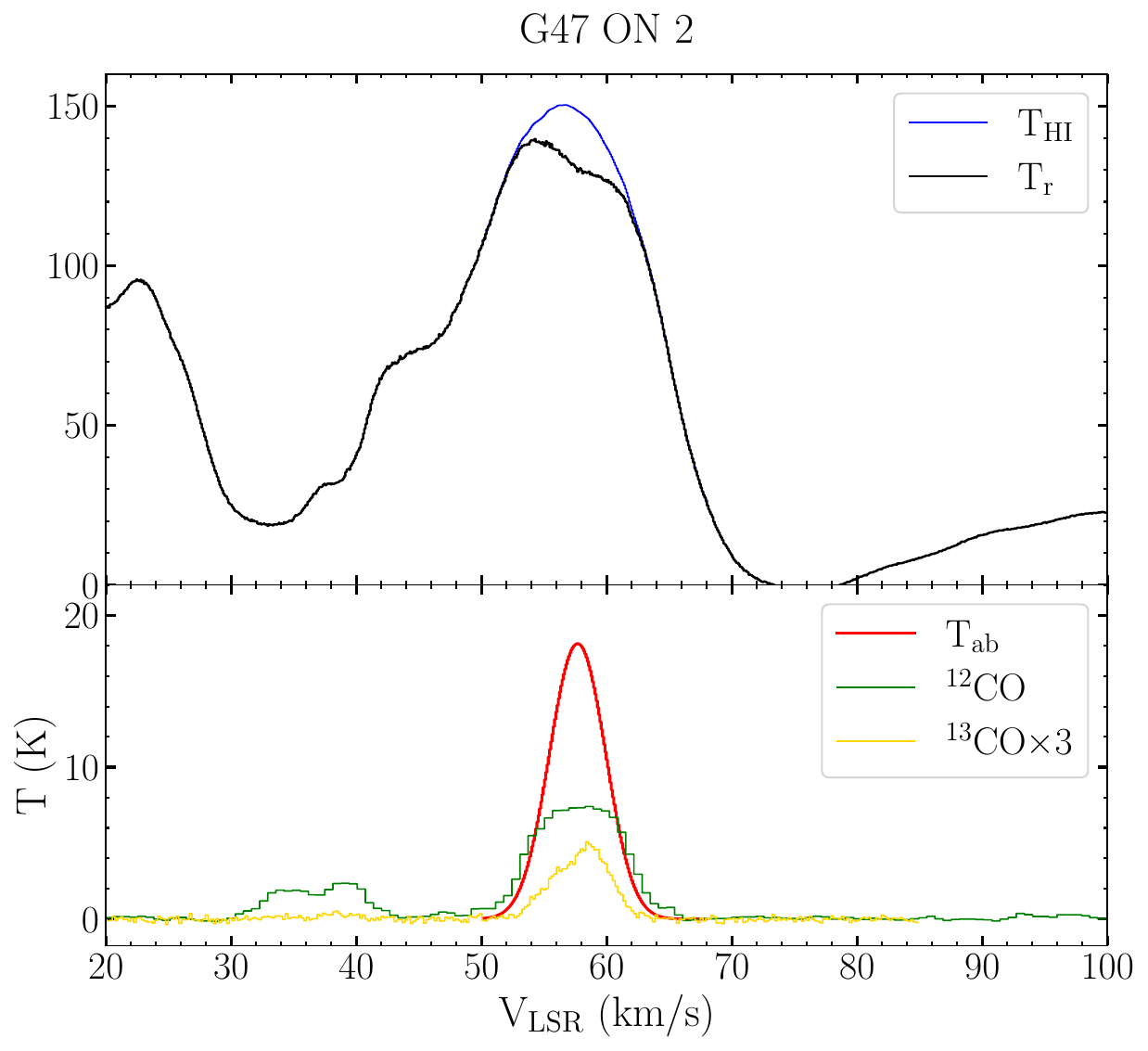}{0.33\textwidth}{(a) ON 2}
          \fig{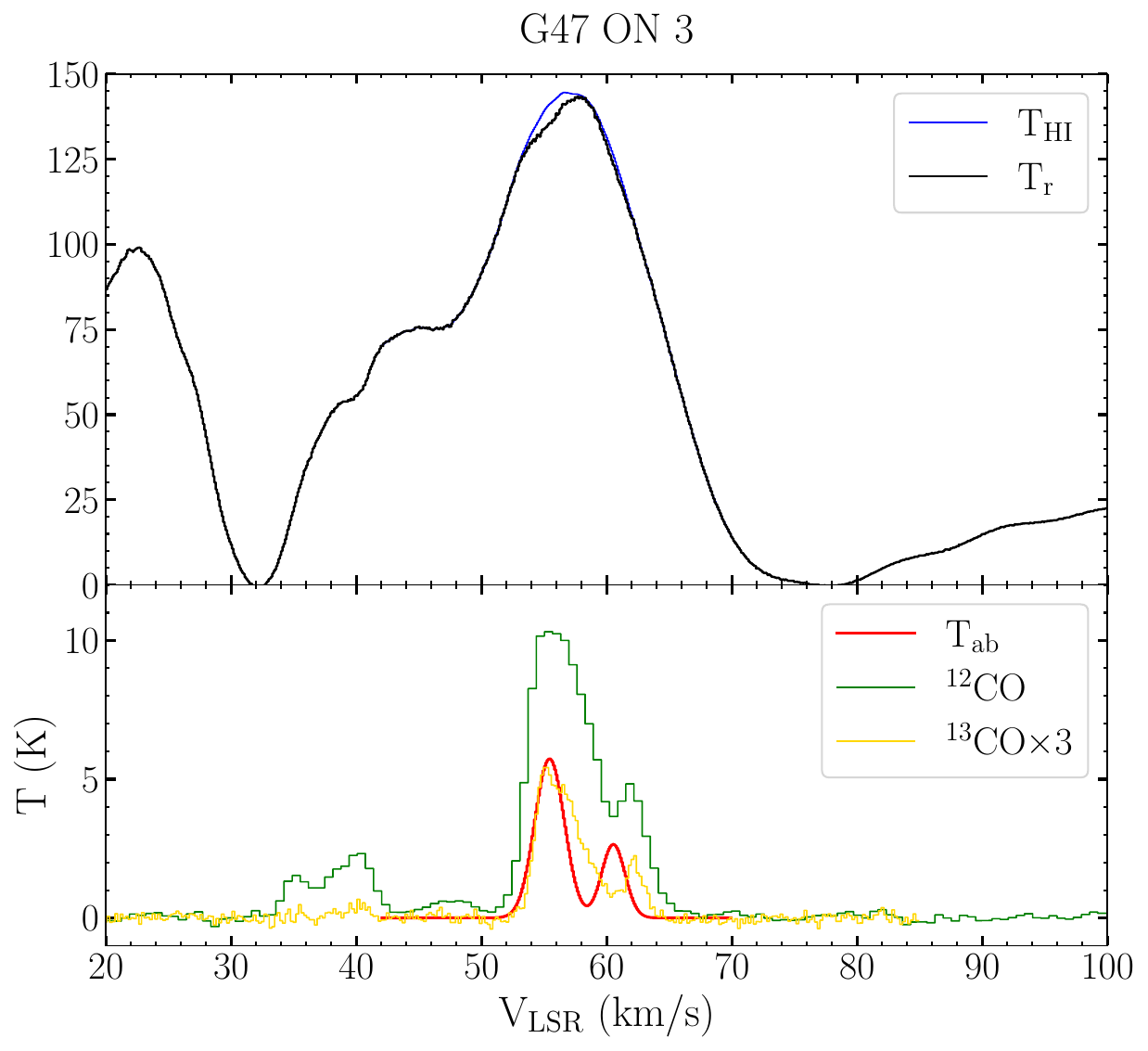}{0.33\textwidth}{(b) ON 3}
          \fig{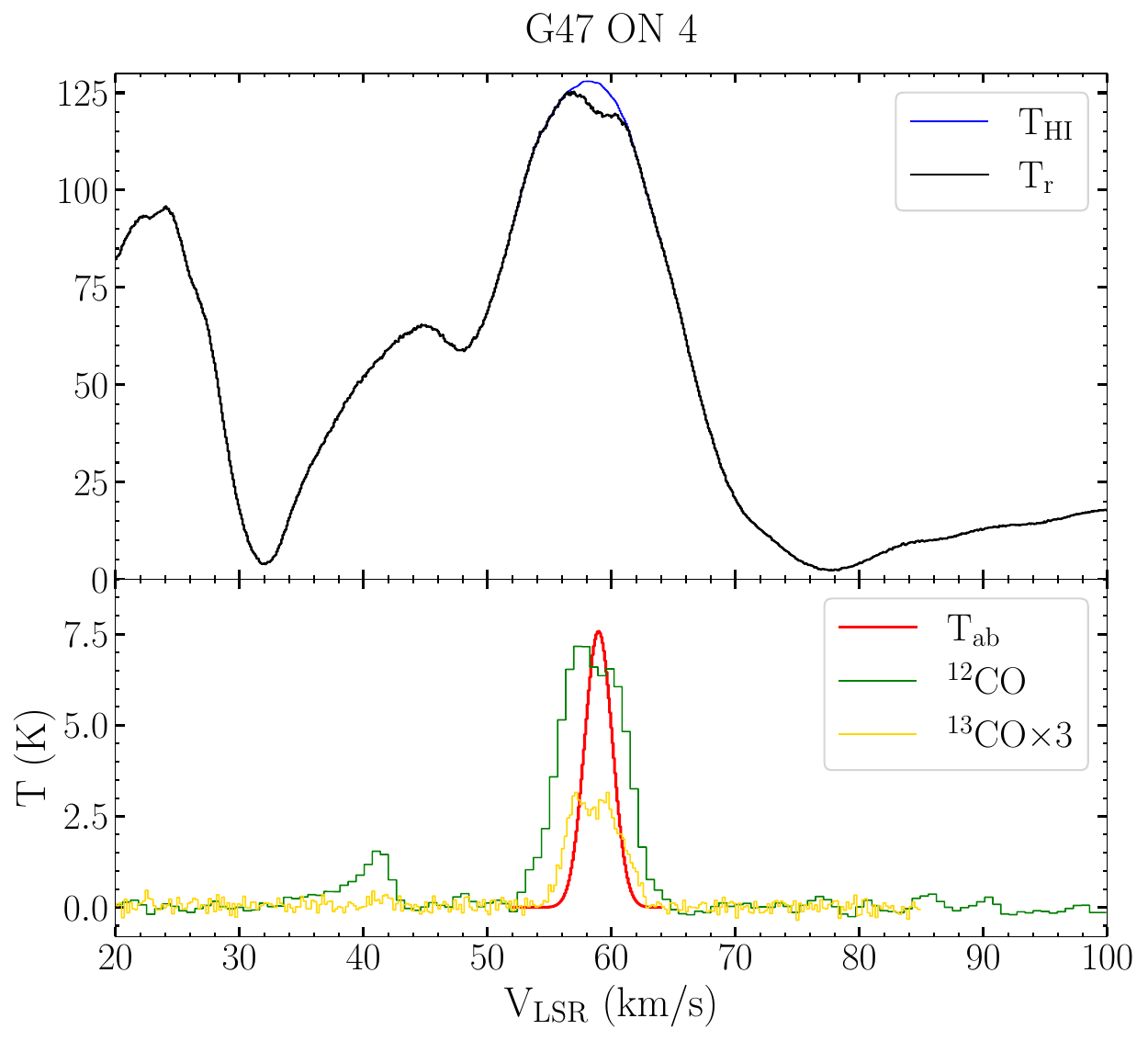}{0.33\textwidth}{(c) ON 4}
          }
\end{figure*}
\begin{figure*}
\gridline{\fig{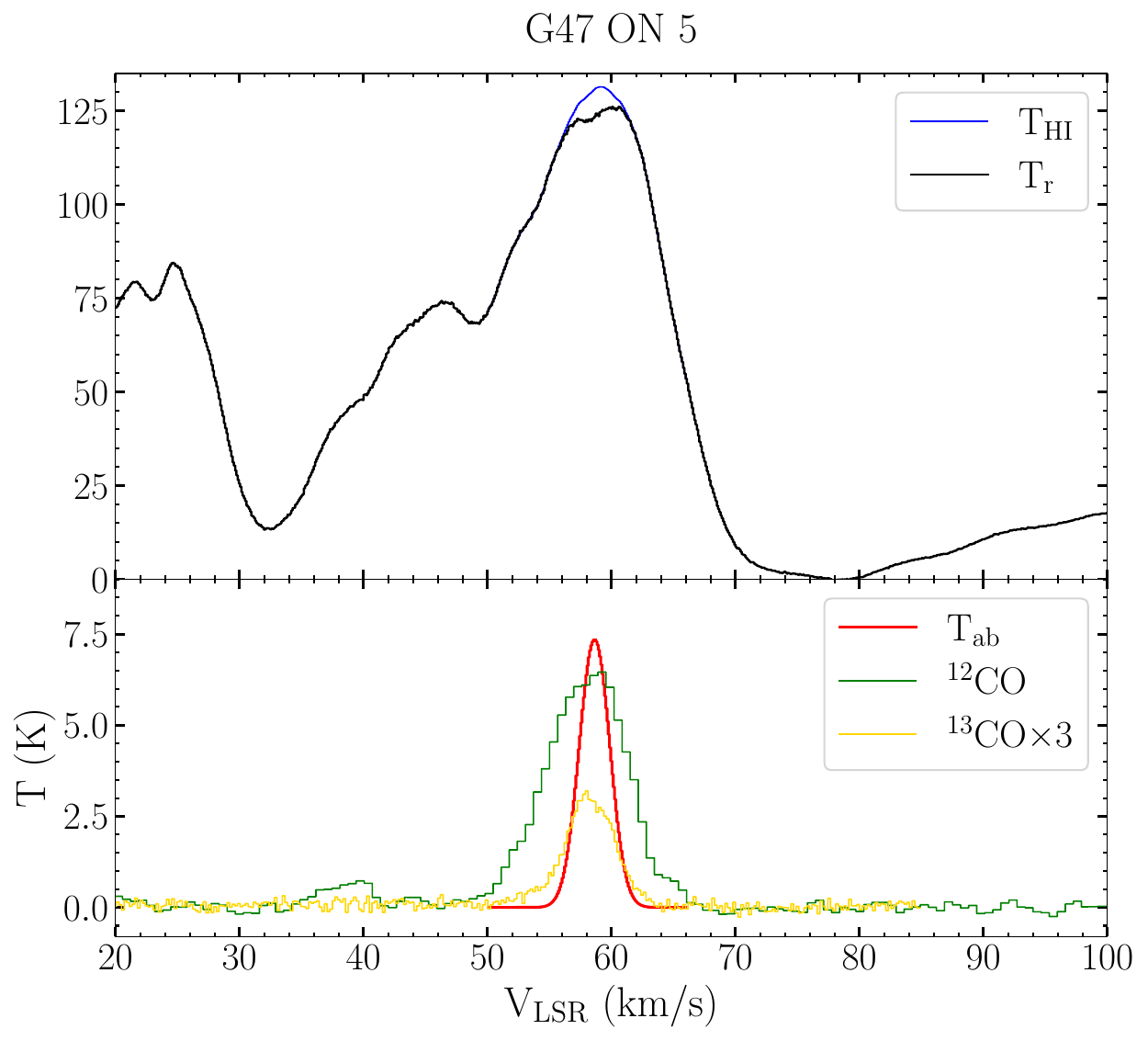}{0.33\textwidth}{(d) ON 5}
		  \fig{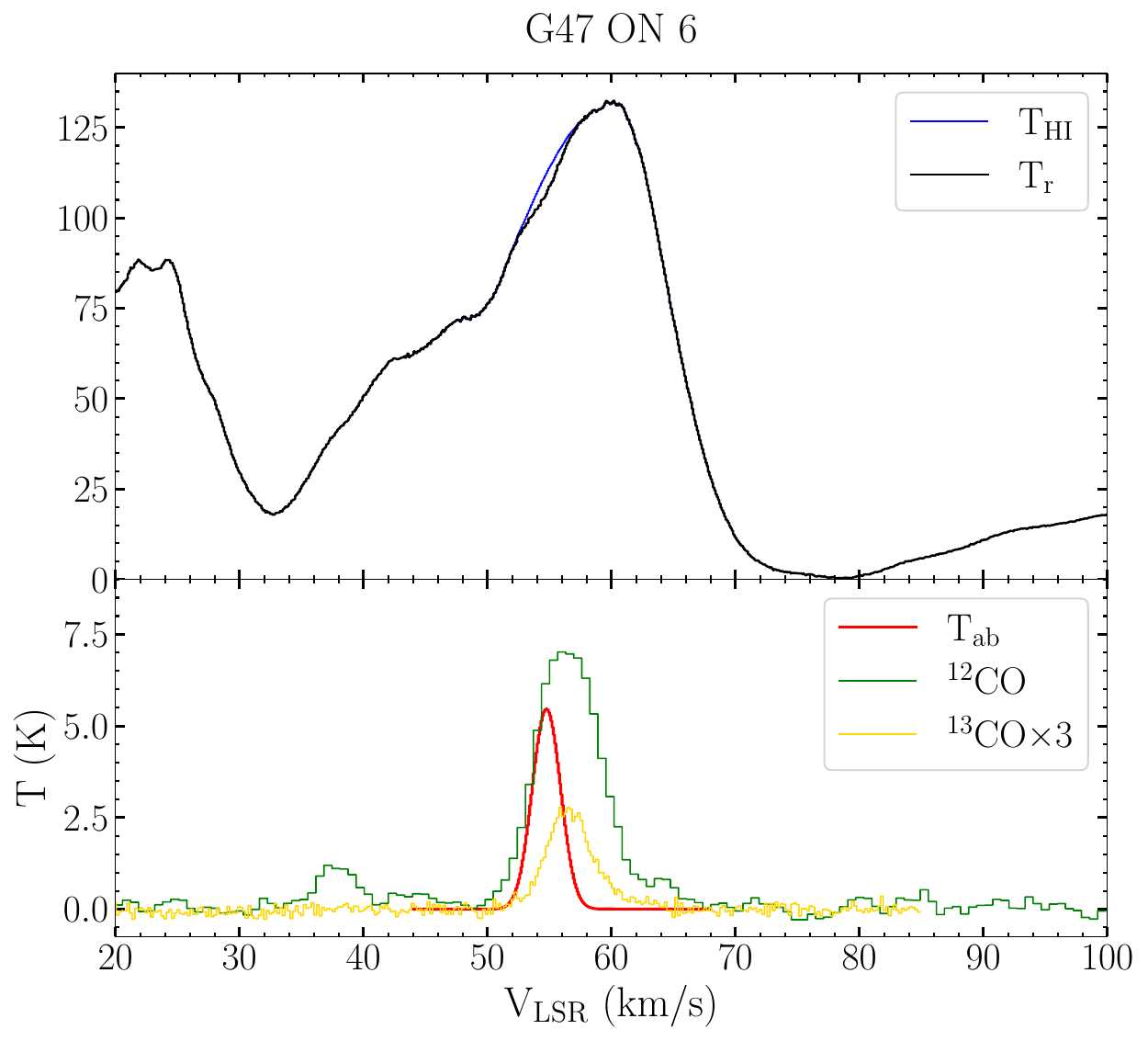}{0.33\textwidth}{(e) ON 6}
		  \fig{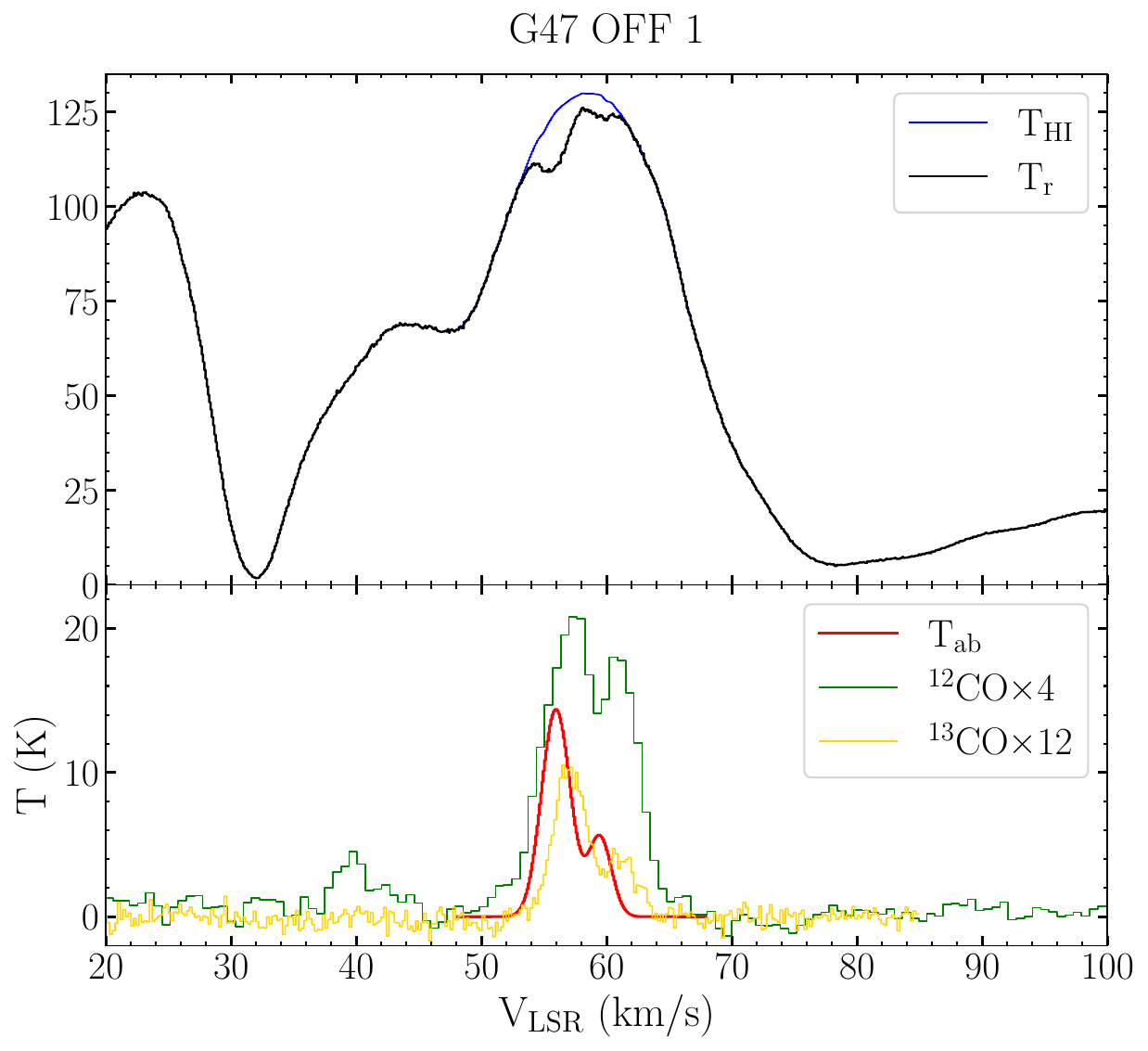}{0.33\textwidth}{(f) OFF 1}
		  }
\gridline{\fig{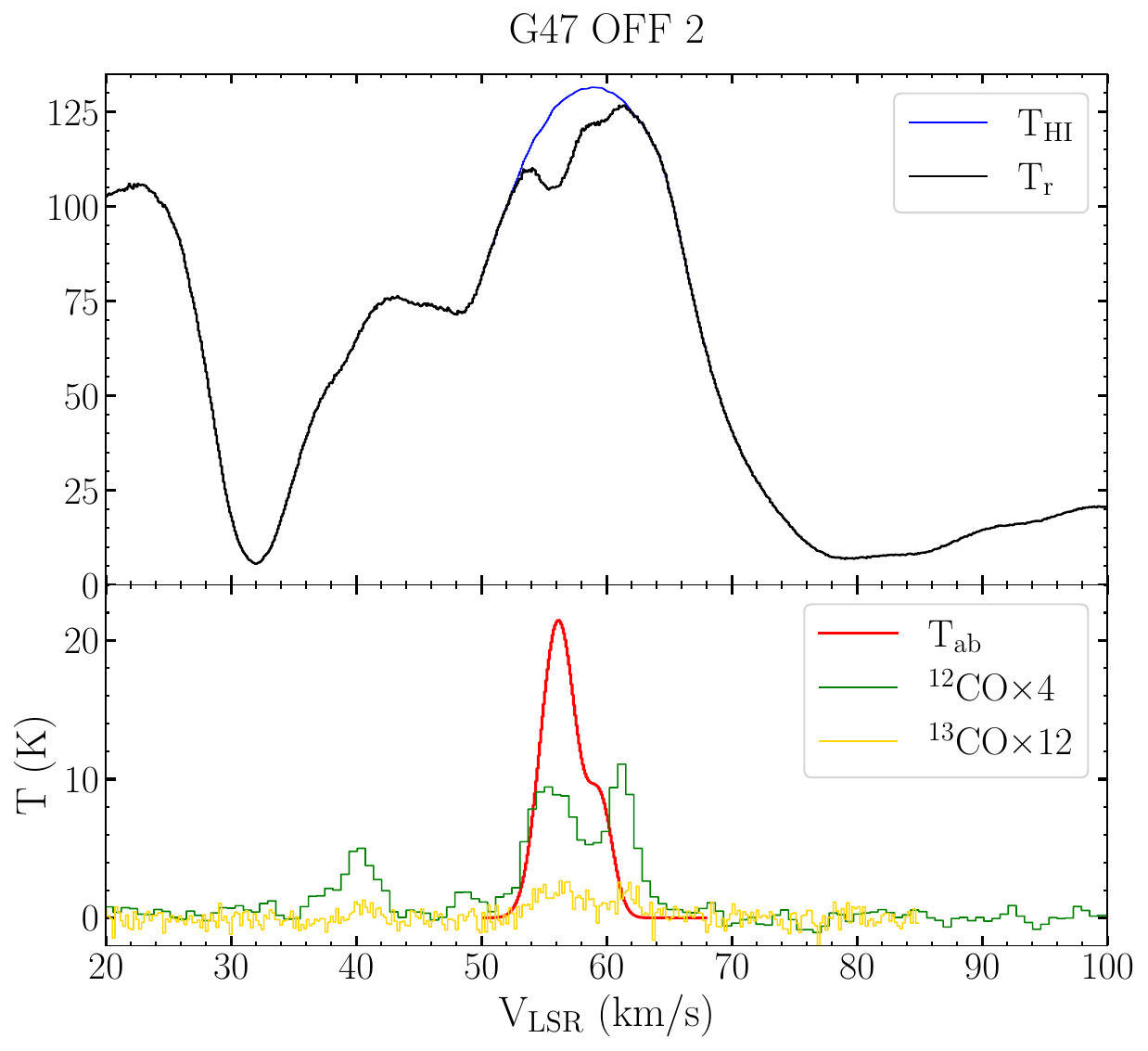}{0.33\textwidth}{(g) OFF 2}
		  \fig{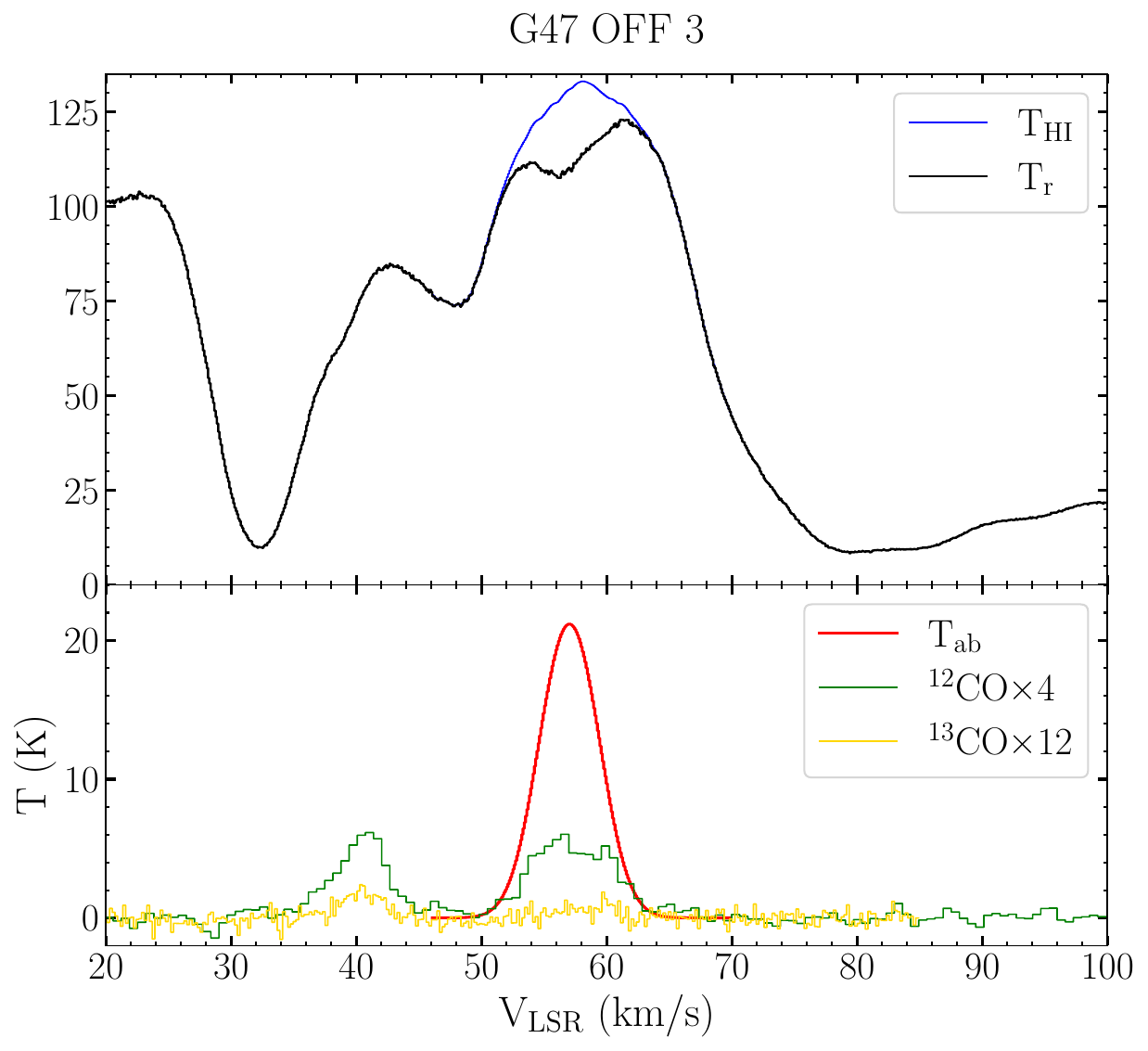}{0.33\textwidth}{(h) OFF 3}
		  \fig{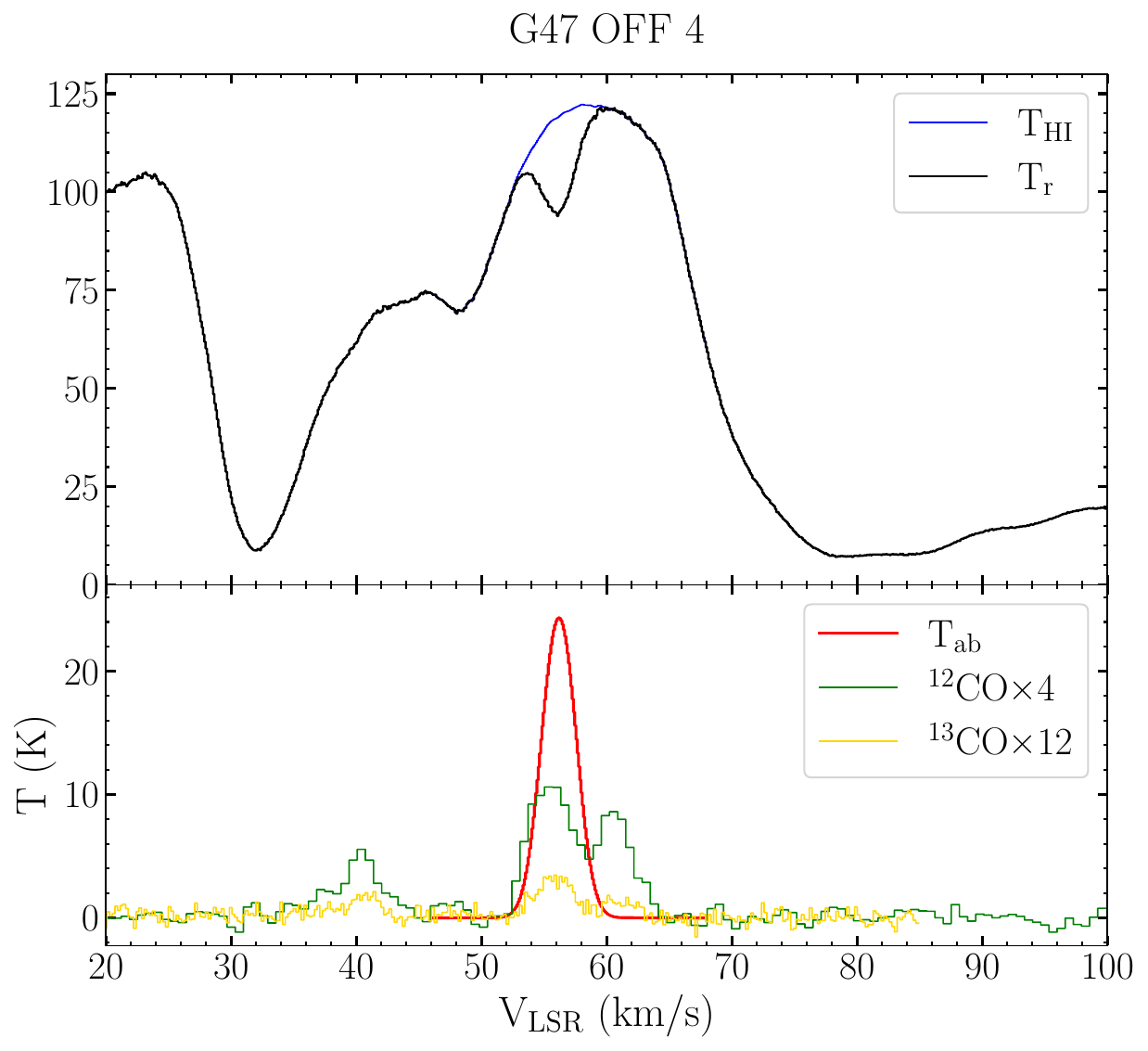}{0.33\textwidth}{(i) OFF 4}
		  }
\end{figure*}
\begin{figure*}
\gridline{\fig{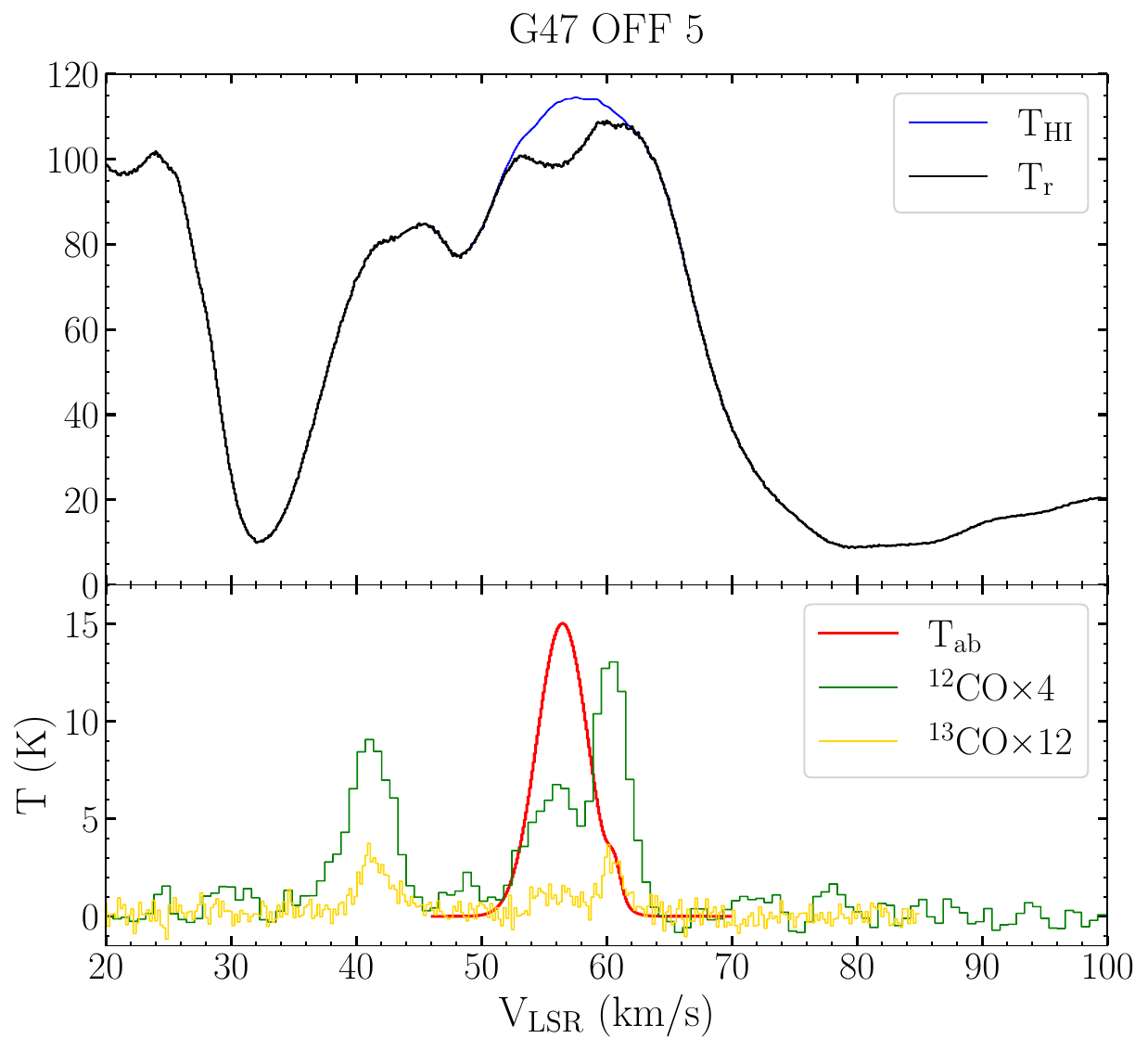}{0.33\textwidth}{(j) OFF 5}
		  \fig{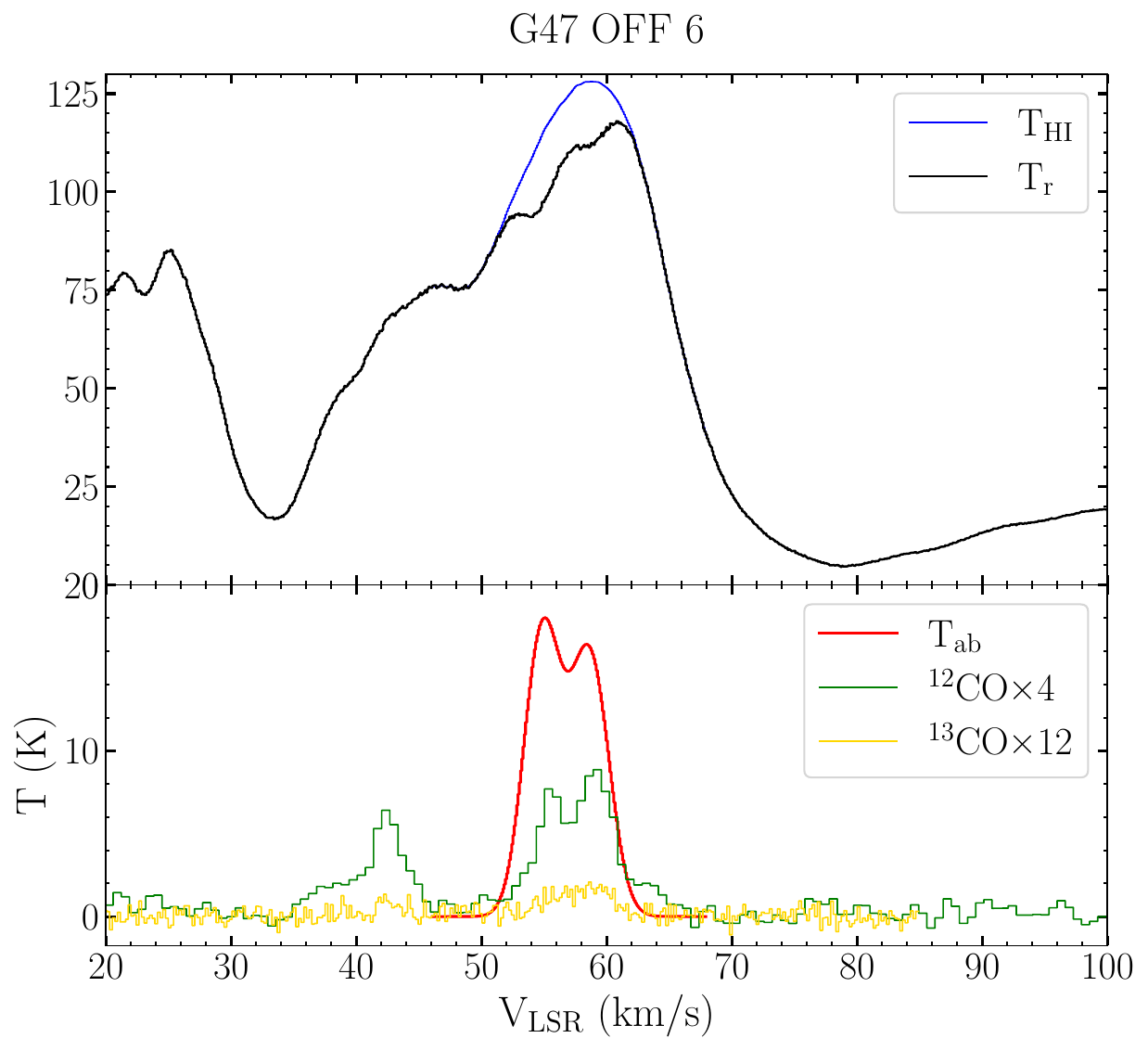}{0.33\textwidth}{(k) OFF 6}
		  }
\caption{HINSA fitting and CO spectra towards G47. Descriptions are the same as Figure\,\ref{fig:4Figs}.}
\label{fig:HINSA G47}
\end{figure*}
\FloatBarrier
\bibliography{sample631}{}
\bibliographystyle{aasjournal}
\end{document}